\documentclass[12pt, a4paper]{article}
\usepackage[dvipdfmx]{graphicx}
\usepackage{amssymb}
\usepackage{amsmath}
\usepackage{bm}
\usepackage{color}
\usepackage{theorem}
\usepackage{subcaption}
\usepackage{listings}
\usepackage{colortbl}
\usepackage{tabularx}
\usepackage{multirow}
\usepackage{longtable}
\usepackage{enumerate}
\usepackage[utf8]{inputenc}
\usepackage[T1]{fontenc}
\usepackage{lmodern}
\usepackage{cancel}
\usepackage[top=30truemm,bottom=30truemm,left=25truemm,right=25truemm]{geometry}

\usepackage[sort&compress,numbers, merge]{natbib}

\definecolor{Orange}{cmyk}{0,0.61,0.87,0}
\definecolor{JungleGreen}{cmyk}{0.99,0,0.52,0}
\definecolor{OliveGreen}{cmyk}{0.64,0,0.95,0.40}
\definecolor{Brown}{cmyk}{0,0.81,1,0.60}
\definecolor{RoyalBlue}{cmyk}{0.71,0.53,0,0.12}
\definecolor{Gray}{cmyk}{0,0,0,0.40}
\definecolor{LightPink}{cmyk}{0.0,0.25,0,0}
\definecolor{LLightPink}{cmyk}{0.0,0.10,0,0}
\definecolor{LightBlue}{cmyk}{0.25,0,0,0}
\definecolor{LightGray}{cmyk}{0,0,0,0.2}

\newcommand{\Slash}[1]{{\ooalign{\hfil/\hfil\crcr$#1$}}}

\newcommand{\vev}[1]{\left\langle #1\right\rangle}

\newcommand{\mrm}[1]{\mathrm{#1}}
\newcommand{\nn}{\nonumber\\}
\newcommand{\bs}[1]{\boldsymbol{#1}}

\allowdisplaybreaks[1]

\usepackage[colorlinks=true, linkcolor=OliveGreen, citecolor=RoyalBlue,
urlcolor=RoyalBlue]{hyperref}

\renewcommand{\thefootnote}{\fnsymbol{footnote}}

\begin{document}

\begin{titlepage}

  \begin{flushright}
    {\tt
    }
\end{flushright}

\vskip 1.35cm
\begin{center}

{\Large
{\bf
Majoron-Driven Leptogenesis in Gauged $U(1)_{L_{\mu}-L_{\tau}}$ Model}
}

\vskip 1.5cm

\renewcommand*{\thefootnote}{\fnsymbol{footnote}}
Juntaro~Wada$^{a~}$\footnote{\href{mailto:wada@hep-th.phys.s.u-tokyo.ac.jp }{\tt wada@hep-th.phys.s.u-tokyo.ac.jp}}

\vskip 0.8cm
{\it ${}^a$Department of Physics, University of Tokyo, Bunkyo-ku, Tokyo 113--0033, Japan,} \\[2pt]

\date{\today}

\vskip 1.5cm

\begin{abstract}
We propose a novel leptogenesis scenario in the gauged $U(1)_{L_{\mu}-L_{\tau}}$ model. 
Achieving successful leptogenesis in the $U(1)_{L_{\mu}-L_{\tau}}$ symmetric phase is challenging due to the absence of a CP phase, caused by restriction from the gauge symmetry. To overcome this issue, we introduce an additional global symmetry, $U(1)_{B-L}$, and a scalar field $\Phi$ responsible for breaking this symmetry. Through the kinetic misalignment mechanism, the majoron field associated with $U(1)_{B-L}$ symmetry breaking has a kinetic motion in the early universe. Subsequently, time-dependent majoron field background induces the background CP phase dynamically, leading to successful leptogenesis in the $U(1)_{L_{\mu}-L_{\tau}}$ symmetric phase. Furthermore, majoron itself serves as a dark matter candidate in this scenario. As one of the phenomenological applications, we consider the model that can also explain the muon $g-2$ anomaly.
\end{abstract}

\end{center}
\end{titlepage}
\renewcommand{\thefootnote}{\arabic{footnote}}
\setcounter{footnote}{0}

\section{Introduction}
\label{sec:intro}

The origin of baryon asymmetry, or matter-antimatter asymmetry, in the current universe, remains a profound puzzle in particle physics and cosmology. Leptogenesis~\cite{Fukugita:1986hr} stands out as a compelling mechanism capable of explaining the observed baryon asymmetry. In this mechanism, right-handed Majorana neutrinos play an important role; their mass breaks the lepton number and causes baryon number breaking through the sphaleron process~\cite{Kuzmin:1985mm}, and their decay to the standard model leptons satisfies the conditions of CP violation and departure from thermal equilibrium, known as Sakharov's conditions~\cite{Sakharov:1967dj}. Furthermore, the seesaw mechanism~\cite{Minkowski:1977sc, Yanagida:1979as, GellMann:1980vs, Mohapatra:1979ia} allows us to address the puzzle of the smallness of neutrino masses~\cite{ParticleDataGroup:2022pth}.

In this paper, our focus is on a model featuring the gauged $U(1)_{L_{\mu}-L_{\tau}}$ symmetry~\cite{Foot:1990mn, He:1990pn, He:1991qd, Foot:1994vd}. Within this framework, $\mu$ flavor particles carry $+1$ charge, while $\tau$ flavor particles carry $-1$ charge. This symmetry strongly influences the neutrino mass structure, because the second and third flavor leptons are charged under this symmetry whereas the first flavor is not. Consequently, the Dirac mass matrix exhibits a simple diagonal structure, and only specific components of the Majorana mass matrix are non-zero. To enable neutrino oscillation, we need to introduce scalars that break this gauged $U(1)_{L_{\mu}-L_{\tau}}$ symmetry~\cite{Branco:1988ex, Choubey:2004hn}. After $U(1)_{L_{\mu}-L_{\tau}}$ symmetry breaking, it remains non-trivial whether the observed neutrino oscillation parameters can be reproduced. In the minimal setup consistent with neutrino oscillations, several predictions for neutrino oscillation data have been obtained ~\cite{Araki:2012ip, Heeck:2014sna, Crivellin:2015lwa, Asai:2017ryy, Asai:2018ocx, Asai:2019ciz}.

While the scale of $U(1)_{L_{\mu}-L_{\tau}}$ symmetry breaking does not significantly impact the consistency with neutrino oscillation data, it holds significance for other phenomena. A high breaking scale is favored in successful leptogenesis~\cite{Asai:2020qax, Granelli:2023egb}, while a low breaking scale is motivated by contexts such as muon $g-2$ anomaly~\cite{Gninenko:2001hx, Baek:2001kca, Murakami:2001cs, Ma:2001md, Ma:2001tb}, Hubble tension~\cite{Escudero:2019gzq, Araki:2021xdk}, and searches for dark sector particles~\cite{Asai:2021xtg, Moroi:2022qwz}.~\footnote{One can introduce the $U(1)_{L_{\mu}-L_{\tau}}$ charged dark matter candidate~\cite{Baek:2008nz, Altmannshofer:2016jzy, Arcadi:2018tly, Bauer:2018egk, Okada:2019sbb}, but we will not discuss this possibility.} Some efforts have been made to achieve successful leptogenesis with a low-scale $U(1)_{L_{\mu}-L_{\tau}}$ breaking scalar~\cite{Borah:2021mri, Eijima:2023yiw}. However, these models lose predictive power for neutrino oscillation due to the need for more additional scalars that break $U(1)_{L_{\mu}-L_{\tau}}$ symmetry. The reason why we need more $U(1)_{L_{\mu}-L_{\tau}}$ breaking scalar is that firstly, the production of sufficient baryon asymmetry typically requires a high temperature which is far from the low $U(1)_{L_{\mu}-L_{\tau}}$ breaking scale.  Then, at such high temperatures, the $U(1)_{L_{\mu}-L_{\tau}}$ symmetry can be restored, potentially leading to a lack of CP phases during leptogenesis due to symmetry restrictions. To circumvent this issue, additional scalars that break $U(1)_{L_{\mu}-L_{\tau}}$ symmetry and their interactions are necessary, compromising predictive power.

In our work, we propose a novel model capable of achieving successful leptogenesis with a low-scale $U(1)_{L_{\mu}-L_{\tau}}$ breaking without sacrificing predictive power for neutrino oscillation data. We introduce an additional symmetry, a global $U(1)_{B-L}$, and a scalar field $\Phi$ responsible for breaking this global symmetry. Suppose this scalar field $\Phi$ has a kinetic motion to the phase direction in the early universe, similar to the Affleck-Dine mechanism~\cite{Affleck:1984fy, Dine:1995kz}. Then a time-dependent background phase is induced, enabling successful leptogenesis in the $U(1)_{L_{\mu}-L_{\tau}}$ symmetric phase. This scenario can be realized through the kinetic misalignment mechanism~\cite{Co:2019jts, Chang:2019tvx, Co:2019wyp}, and the angular component of the rotating field, majoron, can serve as a dark matter candidate. While leptogenesis with time-dependent background phase has been studied in several works~\cite{Kusenko:2014uta, Ibe:2015nfa, Domcke:2020kcp, Co:2020jtv, Domcke:2020quw, Berbig:2023uzs, Chao:2023ojl, Chun:2023eqc, Barnes:2024jap}, we incorporate the phenomenological aspects specific to the gauged $U(1)_{L_{\mu}-L_{\tau}}$ model. As one of the phenomenological applications, we demonstrate that our proposed scenario is applicable in models explaining the muon $g-2$ anomaly.

The paper is organized as follows. In Sec.~\ref{sec: Model}, we present our model and its basic properties, highlighting the challenges for successful leptogenesis in the $U(1)_{L_{\mu}-L_{\tau}}$ symmetric phase. In Sec.~\ref{sec: LG from Majoron}, we briefly explain how to achieve successful leptogenesis in the $U(1)_{L_{\mu}-L_{\tau}}$ symmetric phase. We then evaluate the produced baryon asymmetry and dark matter abundance, discussing various constraints on our model. In Sec.~\ref{sec: result}, we present the parameter space compatible with the observed baryon asymmetry and the dark matter abundance.

\section{Model}
\label{sec: Model}

Our model is constructed upon the framework of the gauged $U(1)_{L_{\mu}-L_{\tau}}$ symmetry~\cite{Foot:1990mn, He:1990pn, He:1991qd, Foot:1994vd}. In this framework, we assign charges to the $+1$ for $\mu$ flavor fields and $-1$ for $\tau$ flavor fields. To reproduce the neutrino mixing and be consistent with the neutrino oscillation data, we introduce three heavy right-handed neutrinos (denoted as $N_e, N_{\mu}$, and $N_{\tau}$) along with the standard model gauge singlet scalars, $\sigma_{\mu}$ and $\sigma_{\tau}$.\footnote{In our model, two scalar fields, $\sigma_{\mu}$ and $\sigma_{\tau}$, are introduced to break the $U(1)_{L_{\mu}-L_{\tau}}$ symmetry. Alternatively, a model consistent with neutrino oscillations can be constructed using a single scalar field $\sigma$~\cite{Harigaya:2013twa, Kaneta:2016uyt, Asai:2017ryy, Asai:2018ocx, Nomura:2019uyz, Asai:2019ciz}. However, in this case, the interaction involving $\sigma^{*}$, which is necessary to reproduce the correct neutrino mixing, explicitly breaks $U(1)_{B-L}$ and typically leads to a large majoron mass. Such a heavy majoron would spoil our scenario due to the overabundance of majoron dark matter.} They carry the $U(1)_{L_{\mu}-L_{\tau}}$ charges $0,+1,-1$, and $+1,-1$ respectively. Our model also contains $U(1)_{L_{\mu}-L_{\tau}}$ gauge boson $Z'$.

Furthermore, our model incorporates an additional symmetry, namely {\it global} $U(1)_{B-L}$, and a scalar field $\Phi$ responsible for breaking this global symmetry. Under this symmetry, we assign charges to the $-1$ for left-handed $SU(2)$ lepton doublets ($\ell_{e}, \ell_{\mu}$, and $\ell_{\tau}$), the right-handed charged leptons ($e_R, \mu_R$, and $\tau_R$) and  three heavy right-handed neutrinos ($N_e,N_{\mu}$, and $N_{\tau}$), while left-handed $SU(2)$ quark doublets $Q$ and right-handed charged quarks $u_R,d_R$ carry a $U(1)_{B-L}$ charge of $+1/3$. The singlet scalars, $\sigma_{\mu}$, $\sigma_{\tau}$ and $\Phi$, have $U(1)_{B-L}$ charge of $-2$.

We summarise these charges assignments of the field content in Table~\ref{tab: model}.

\newcolumntype{C}{@{}>{\centering\arraybackslash}X}
\begin{table}[h!]
 \begin{tabularx}{\textwidth}{c|C|C|C|C|C|c|c}
\hline
\rowcolor[gray]{.95}
    \multicolumn{8}{c}{\bf Charges and Field Content}\\
    \hline
    \hline
  & $\ell_{e}, e_R, N_e$ & $\ell_{\mu}, \mu_R, N_\mu$ & $\ell_{\tau }, \tau_R, N_\tau$ &  $\sigma_{\mu}$,~$\sigma_{\tau}$  & $\Phi$ &$Q,u_R,d_R$ & Others  \\
 \hline
 ${L_\mu - L_\tau}$& $0$ &  $+1$ & $-1$ & $+1$,~$-1$ & $0$ & $0$ & $0$ \\
 \hline
 ${B-L}$& $-1$ &  $-1$ & $-1$ & $-2$ & $-2$ & $+1/3$ & $0$\\
 \hline
 \hline
 \end{tabularx}
 \caption{Assigned charges of the field content in our model.}
   \label{tab: model}
 \end{table}
Under this setup, we can write down the Lagrangian for the neutrino sector as follows:\footnote{We adopt a two-component spinor notation (see, e.g., Ref.~\cite{Dreiner:2008tw}) as in Ref.~\cite{Asai:2017ryy}.}
\begin{align}
{\cal L}_{N} =
&-\lambda_e N_e^c (\ell_{e} \cdot H)
-\lambda_\mu N_\mu^c (\ell_{\mu} \cdot H)
-\lambda_\tau N_\tau^c (\ell_{\tau} \cdot H) \nonumber \\
&-\frac{1}{2}\lambda_{ee}\Phi N_e^c N_e^c 
- \lambda_{\mu\tau}\Phi N_\mu^c N_\tau^c 
- \lambda_{e\mu} \sigma_{\mu} N_e^c N_\mu^c
- \lambda_{e\tau} \sigma_{\tau} N_e^c N_\tau^c +\text{h.c.} ~,
\label{eq:Lagrangian}
\end{align}
where  $H$ is the Standard Model Higgs boson, and the dots indicate the contraction of the SU(2) indices between the lepton doublets and the Higgs doublet.

After the Higgs field $H$, and singlets $\sigma_{\mu}$, $\sigma_{\tau}$ and $\Phi$ get their vacuum expectation value (VEVs),\footnote{
We assume that the scalar potential can be expressed as $V(H,\sigma_{\mu},\sigma_{\tau},\Phi)=V(H)+V(\sigma_{\mu})+V(\sigma_{\tau})+V(\Phi)$. We will present the explicit form of the scalar potential $V(\Phi)$ that we considered in Sec.~\ref{subsection: evaluation}.
}  denoted as $\vev{H}=v/\sqrt{2}$, $\vev{\sigma_{\mu}}$, $\vev{\sigma_{\tau}}$ and $\vev{\Phi}$, these interaction terms lead to the neutrino mass terms:
\begin{align}
 {\cal L}_{\rm mass} &= -(\nu_{e}, \nu_{\mu}, \nu_{\tau}) {\cal M}_D 
\begin{pmatrix}
N_e^c\\ N_\mu^c\\ N_\tau^c 
\end{pmatrix}
- \frac{1}{2}(N_e^c, N_\mu^c, N_\tau^c) {\cal M}_R 
\begin{pmatrix}
 N_e^c \\ N_\mu^c \\ N_\tau^c 
\end{pmatrix}
+\text{h.c.} ~,
\end{align}
where $\nu_\alpha$ are the SM neutrinos of lepton flavour $\alpha$, ${\cal M}_D$ is the Dirac mass matrix and $ {\cal M}_R$ is the Majorana mass matrix given by,
\begin{equation}
\label{eq: Dirac and Majorana mass}
 {\cal M}_D = \frac{v}{\sqrt{2}}
\begin{pmatrix}
 \lambda_e & 0& 0\\
 0 & \lambda_\mu & 0 \\
 0 & 0 & \lambda_\tau 
\end{pmatrix}
~,\qquad
{\cal M}_R =
\begin{pmatrix}
 \lambda_{ee}\vev{\Phi} & \lambda_{e\mu} \vev{\sigma_{\mu}} & \lambda_{e\tau} \vev{\sigma_{\tau}} \\
 \lambda_{e\mu} \vev{\sigma_{\mu}} & 0 & \lambda_{\mu\tau}\vev{\Phi} \\
\lambda_{e\tau} \vev{\sigma_{\tau}} & \lambda_{\mu\tau}\vev{\Phi} & 0
\end{pmatrix}
~,
\end{equation}
respectively. We can simplify our analysis by choosing the $U(1)_{L_{\mu}-L_{\tau}}$ breaking VEVs, $\vev{\sigma_{\mu}}$ and $\vev{\sigma_{\tau}}$, to be real and positive through field redefinitions without loss of generality. Additionally, we can make the Dirac Yukawa couplings $\lambda_e$, $\lambda_{\mu}$, and $\lambda_{\tau}$ real and positive. These Yukawa couplings can be parameterized as:
\begin{equation}
  (\lambda_e, \lambda_\mu, \lambda_\tau) = \lambda (\cos \theta, \sin \theta \cos \phi, \sin \theta \sin \phi),
  \label{eq:lthetaphi}
\end{equation}
where $0 < \lambda \leq 1$, and $0 \leq \theta, \phi \leq \pi/2$.

The light neutrino mass matrix is obtained from the Type-I seesaw formula~\cite{Minkowski:1977sc, Yanagida:1979as, GellMann:1980vs, Mohapatra:1979ia}, 
\begin{align}
{\cal M}_\nu = -{\cal M}_D {\cal M}_{R}^{-1}{\cal M}^{T}_D.
\label{eq: seesaw}
\end{align}
Since it is a complex symmetric matrix, it can be diagonalized with a unitary matrix as $U^T {\cal M}_{\nu} U =\text{diag}(m_1, m_2, m_3)$, where the unitary matrix $U$ is the Pontecorvo-Maki-Nakagawa-Sakata (PMNS) neutrino mixing matrix and is given by
\begin{equation}
 U = 
\begin{pmatrix}
 c_{12} c_{13} & s_{12} c_{13} & s_{13} e^{-i\delta} \\
 -s_{12} c_{23} -c_{12} s_{23} s_{13} e^{i\delta}
& c_{12} c_{23} -s_{12} s_{23} s_{13} e^{i\delta}
& s_{23} c_{13}\\
s_{12} s_{23} -c_{12} c_{23} s_{13} e^{i\delta}
& -c_{12} s_{23} -s_{12} c_{23} s_{13} e^{i\delta}
& c_{23} c_{13}
\end{pmatrix}
\begin{pmatrix}
 1 & & \\
 & e^{i\frac{\alpha_{2}}{2}} & \\
 & & e^{i\frac{\alpha_{3}}{2}}
\end{pmatrix}
~,
\end{equation}
where $c_{ij} \equiv \cos \theta_{ij}$ and $s_{ij} \equiv \sin
\theta_{ij}$ with the mixing angles $\theta_{ij}$, and we denote the Dirac CP phase and Majorana CP phases as $\delta$, $\alpha_2$, and $\alpha_3$, respectively. 

Diagonalizing the Majorana mass matrix, we obtain the masses of right-handed neutrinos,
\begin{align}
{\cal M}_R &= \Omega^{\ast} \text{diag}(M_{1},M_{2},M_{3}) \Omega^{\dagger}~, 
\end{align}
where $\Omega$ is a unitary matrix and $M_{1,\,2,\,3}$ are the right-handed neutrino mass and they satisfy $0 < M_1 < M_{2} < M_{3}$.

\begin{figure}[t!]
    \centering
    \includegraphics[width = .48\textwidth]{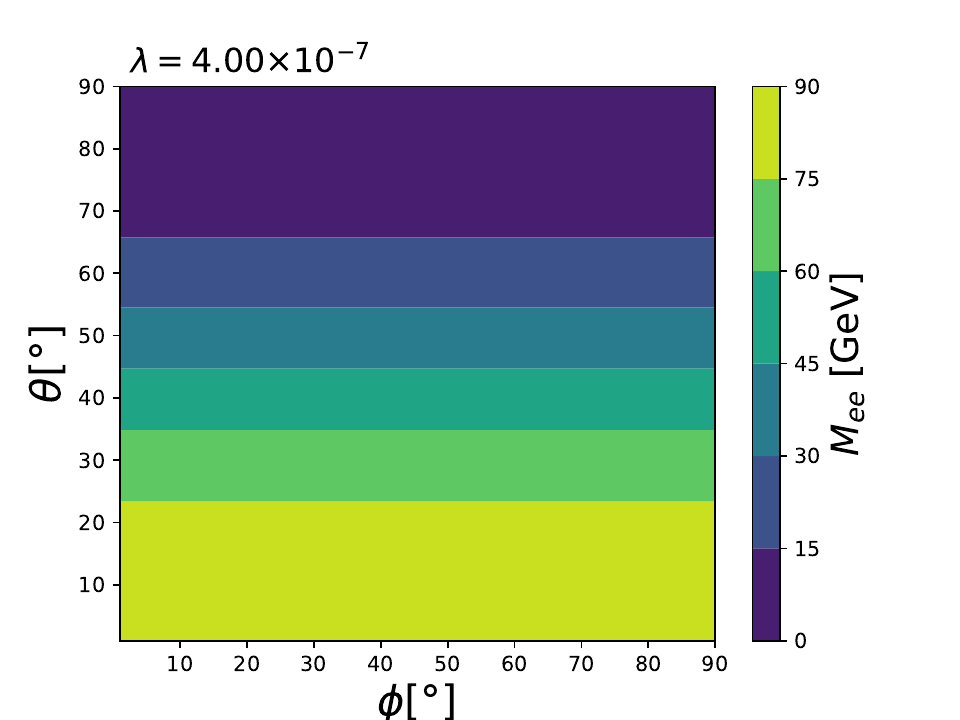}
    \includegraphics[width = .48\textwidth]{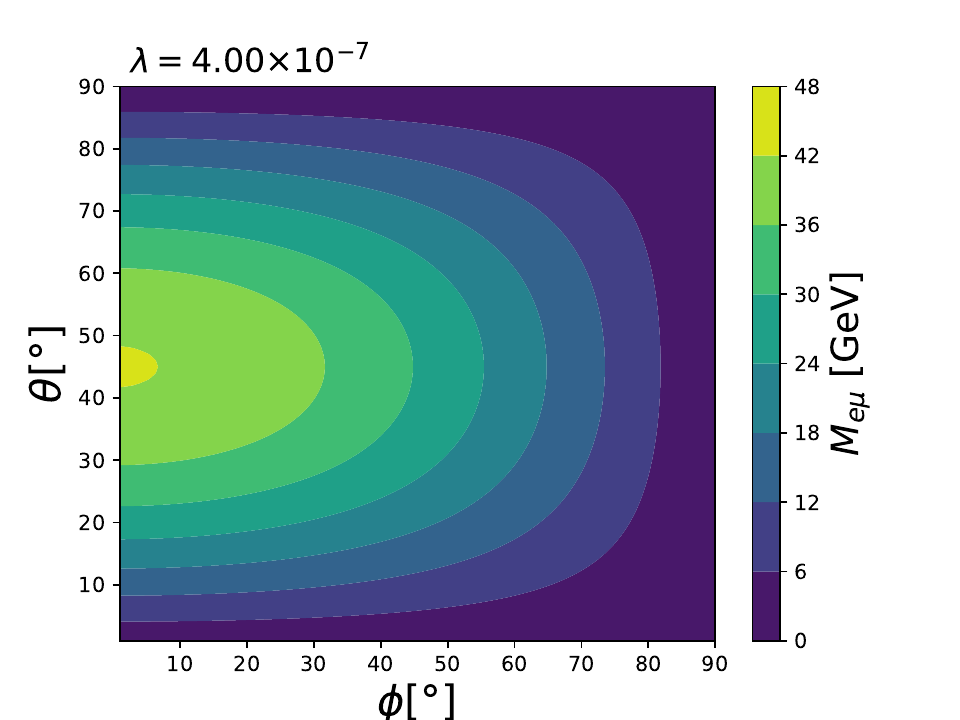}
    \includegraphics[width = .48\textwidth]{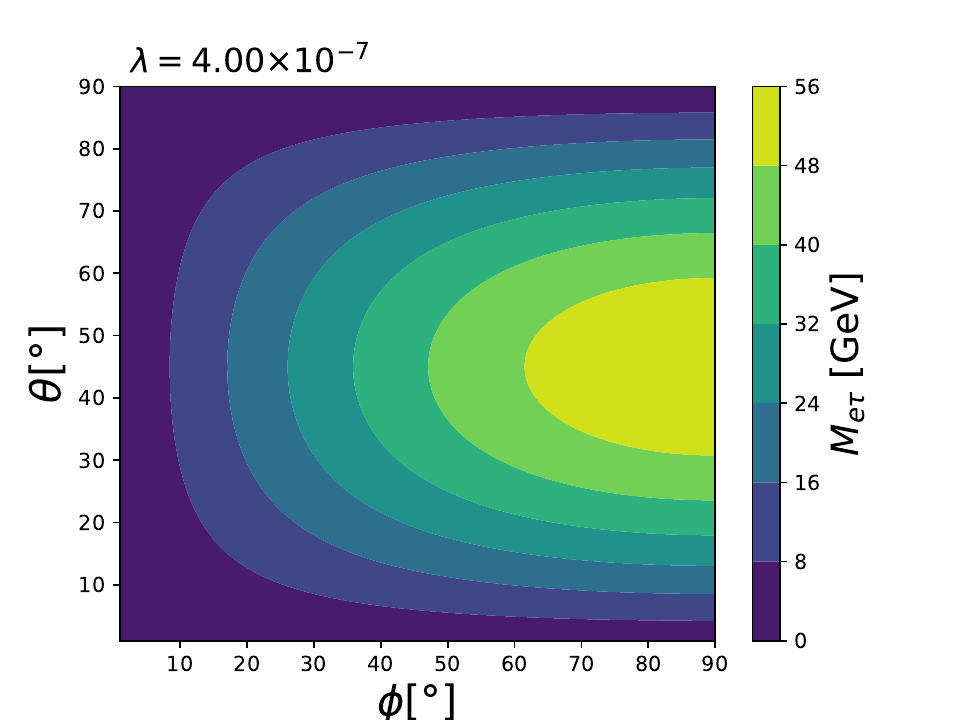}
    \includegraphics[width = .48\textwidth]{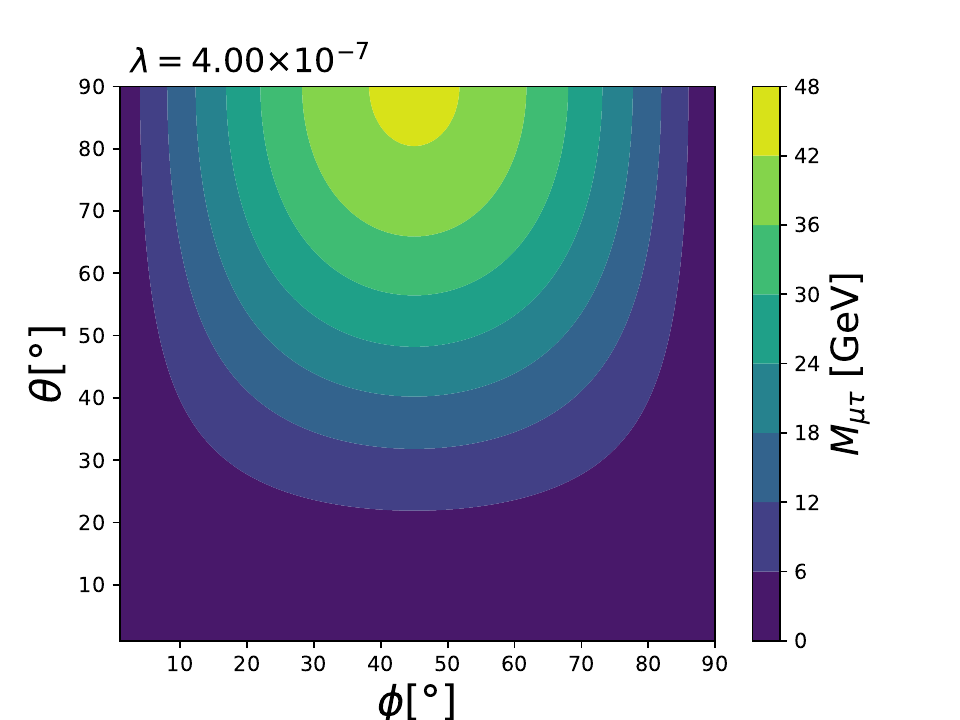}
    \caption{Contours in the $\phi-\theta$ plane illustrating the dependence of each component of the Majorana mass matrix on the model parameters. The plots depict $M_{ee}:=\lambda_{ee}\vev{\Phi}$ (top left), $M_{e\mu}:=\lambda_{e\mu}\vev{\sigma_{\mu}}$ (top right), $M_{e\tau}:=\lambda_{e\tau}\vev{\sigma_{\tau}}$ (bottom left), and $M_{\mu\tau}:=\lambda_{\mu\tau}\vev{\Phi}$ (bottom right). Here, we choose $\lambda = 4 \times 10^{-7}$ for subsequent discussions, and utilize neutrino oscillation parameters from the \texttt{NuFit 5.2} analysis~\cite{nufit, Esteban:2020cvm} (without SK data). Note that relationships derived from Eq.~\eqref{eq: seesaw inverse} reveal that $M_{ee} \propto \lambda^2 \cos^2 \theta$, $M_{e\mu} \propto \lambda^2 \sin \theta \cos \theta \cos \phi$, $M_{e\tau} \propto \lambda^2 \sin \theta \cos \theta \sin \phi$, and $M_{\mu\tau} \propto \lambda^2 \sin^2 \theta \sin \phi \cos \phi$.}
    \label{fig: Majorana mass scan}
\end{figure}

We note that in our model, the phases of the PMNS neutrino mixing matrix and the sum of the standard model neutrino masses are predictable by other neutrino oscillation parameters, because of restricted neutrino mass matrix structure in Eq.~\eqref{eq: Dirac and Majorana mass}, so-called two-zero minor structure~\cite{Araki:2012ip, Heeck:2014sna, Crivellin:2015lwa, Asai:2017ryy, Asai:2018ocx, Asai:2019ciz}. (This structure also appear in the $U(1)_{L_{\mu}-L_{\tau}}$ model with one singlet scalar, called the minimal gauged $U(1)_{L_{\mu}-L_{\tau}}$ model~\cite{Harigaya:2013twa, Kaneta:2016uyt, Asai:2017ryy, Asai:2018ocx, Nomura:2019uyz, Asai:2019ciz}.) Consequently, by fixing the three Dirac Yukawa couplings, we can determine each component of the Majorana mass matrix through the seesaw mechanism formula~\eqref{eq: seesaw}:
\begin{align}
\label{eq: seesaw inverse}
{\cal M}_R = -{\cal M}^{T}_D {\cal M}_{\nu}^{-1} {\cal M}_D.
\end{align}
Here, ${\cal M}_{\nu}$ is entirely determined by observed neutrino oscillation parameters. This implies that we have only three free input parameters $\lambda$, $\theta$, and $\phi$ in the neutrino sector to establish the mass spectrum of right-handed neutrinos and their interactions with standard model particles. As an illustrative example, we present a contour plot (Fig~\ref{fig: Majorana mass scan}) in the $\phi-\theta$ plane for each component of the Majorana mass matrix, denoted as $M_{ee}:=\lambda_{ee}\vev{\Phi}$, $M_{e\mu}:=\lambda_{e\mu}\vev{\sigma_{\mu}}$, $M_{e\tau}:=\lambda_{e\tau}\vev{\sigma_{\tau}}$, and $M_{\mu\tau}:=\lambda_{\mu\tau}\vev{\Phi}$, while fixing $\lambda$. 
We note that that relationships derived from Eq.~\eqref{eq: seesaw inverse} reveal that $M_{ee} \propto \lambda^2 \cos^2 \theta$, $M_{e\mu} \propto \lambda^2 \sin \theta \cos \theta \cos \phi$, $M_{e\tau} \propto \lambda^2 \sin \theta \cos \theta \sin \phi$, and $M_{\mu\tau} \propto \lambda^2 \sin^2 \theta \sin \phi \cos \phi$.

This analysis utilizes neutrino oscillation data from the \texttt{NuFit} analyses \cite{nufit}, which are comprehensive global fits that include data from all the relevant neutrino oscillation experiments. The most recent \texttt{NuFit} analyses (\texttt{NuFit 5.2}~\cite{nufit, Esteban:2020cvm}) are conducted both with and without incorporating data from the Super-Kamiokande (SK) experiment. While these approaches yield different values for the neutrino mixing angles and squared mass differences, in Fig~\ref{fig: Majorana mass scan}, we use without SK data as a representative to determine the light neutrino mass matrix ${\cal M}_{\nu}$.  We note that, to circumvent the constraint on the sum of light neutrino masses, $\sum_{i} m_{i} < (0.12-0.69)~\mathrm{eV}$~\cite{LVinZyla:2020zbs, Capozzi_2020} (see also Refs.~\cite{Vagnozzi:2017ovm, Planck:2018vyg, RoyChoudhury:2019hls, Ivanov:2019hqk, DES:2021wwk, Tanseri:2022zfe}), we set $\theta_{23}$ within the $+3\sigma$ range (detailed in Ref.~\cite{Granelli:2023egb}).

In this study, we focus on the scenario where the breaking scale of $U(1)_{L_{\mu}-L_{\tau}}$ is lower than that of the electroweak phase transition, or more precisely, the temperature at which sphaleron processes freeze out. We make the assumption that the vacuum expectation value (VEV) of the $U(1)_{B-L}$ breaking scalar field, $\Phi$, is significantly larger than the Higgs VEV:
\begin{align}
\vev{\sigma_{\mu}}, \vev{\sigma_{\tau}}  < v/\sqrt{2} \ll \vev{\Phi},
\end{align}
Such a low $U(1)_{L_{\mu}-L_{\tau}}$ breaking scale gives rise to a light $U(1)_{L_{\mu}-L_{\tau}}$ gauge boson since its mass is determined by the $U(1)_{L_{\mu}-L_{\tau}}$ breaking VEV: $m_{Z'} = 2 g' \vev{\sigma}$. Here and in what follows, we denote the $U(1)_{L_{\mu}-L_{\tau}}$ breaking scale as $\vev{\sigma}:=\sqrt{(\vev{\sigma_{\mu}}^2+\vev{\sigma_{\tau}}^2)/2}$, and refer to the mass and coupling of $Z'$ as $m_{Z'}$ and $g'$, respectively. The light $U(1)_{L_{\mu}-L_{\tau}}$ gauge boson is motivated by various phenomenology including the muon $g-2$ anomaly~\cite{Gninenko:2001hx, Baek:2001kca,Murakami:2001cs, Ma:2001md,Ma:2001tb}, Hubble tension~\cite{Escudero:2019gzq,Araki:2021xdk}, and the search for dark sector particles~\cite{Asai:2021xtg,Moroi:2022qwz}. Particularly in the region favored by the muon $g-2$ anomaly, the expected $U(1)_{L_{\mu}-L_{\tau}}$ breaking scale is $\vev{\sigma} = 10- 50~\mathrm{GeV}$~\cite{NA64:2024klw}.\footnote{Several experiments have already ruled out specific regions within the parameter space where the gauged $U(1)_{L_{\mu}-L_{\tau}}$ model could potentially explain the observed muon $g-2$ anomaly~\cite{BaBar:2016sci, CMS:2018yxg, Bellini:2011rx, CHARM-II:1990dvf, CCFR:1991lpl}. More recently, the NA64$\mu$ experiment presented the results, further constraining the remaining ($m_{Z'},g_{Z'}$) parameter space~\cite{NA64:2024klw}. By combining these experimental constraints, we identify a small allowed region characterized by $m_{Z'} \sim 1-4 \times 10^{-2} ~\mathrm{GeV}$ and $g'\sim 4-5\times 10^{-4}$, leading to $\vev{\sigma} = 10- 50~\mathrm{GeV}$.} 

As illustrated in Fig~\ref{fig: Majorana mass scan}, within the central region on the $\phi-\theta$ plane, each component of the Majorana mass matrix exhibits comparable values. This implies a hierarchy among the Yukawa couplings in the Majorana mass matrix, given by:
\begin{align}
\label{eq: coupling hierarchy}
\lambda_{ee}, \lambda_{\mu\tau} \ll \lambda_{e\mu}, \lambda_{e\tau}.
\end{align}
Hereafter, we make the assumption that the Yukawa couplings associated with the $U(1)_{L_{\mu}-L_{\tau}}$ breaking scalar are not significantly small, namely $\lambda_{e\mu}, \lambda_{e\tau} \sim \mathcal{O}(0.1-1)$, while the couplings with the $U(1)_{B-L}$ breaking scalar are substantially smaller. 

In our model, as elucidated thus far, each components in the Majorana mass term are parameterized by three parameters, $(\lambda, \theta, \phi)$ and their magnitudes scale with $\lambda^2$. Therefore, once we fix $(\lambda, \theta, \phi)$, it determines $M_{e\mu}=\lambda_{e\mu}\vev{\sigma_{\mu}}$ and $M_{e\tau}=\lambda_{e\tau}\vev{\sigma_{\tau}}$ components related to the $U(1)_{L_{\mu}-L_{\tau}}$ breaking scale and vice versa. Indeed, the breaking scale favored by the muon $g-2$ anomaly yields a range of $\lambda$, with $\lambda \lesssim 5 \times 10^{-7}$ in the central region of the $\phi-\theta$ plane.

At high temperatures, symmetry broken by scalar VEVs could be restored due to thermal effects and finite density effects~\cite{Eijima:2023yiw}. In the $U(1)_{L_{\mu}-L_{\tau}}$ symmetric phase, the Majorana mass matrix takes the form:
\begin{equation}
\label{eq: Majorana mass matrix in symmetric phase}
{\cal M}_{R,\mrm{sym}} =
\begin{pmatrix}
 \lambda_{ee}\vev{\Phi} & 0 & 0 \\
 0 & 0 & \lambda_{\mu\tau}\vev{\Phi} \\
0 & \lambda_{\mu\tau}\vev{\Phi} & 0
\end{pmatrix}
~,
\end{equation}
where ${\cal M}_{R,\mrm{sym}}$ indicates the Majorana mass matrix in the $U(1)_{L_{\mu}-L_{\tau}}$ symmetric phase.
Since most of the components in the neutrino mass matrix are restricted from $U(1)_{L_{\mu}-L_{\tau}}$ symmetry, we lack sufficient physical phases in the $U(1)_{L_{\mu}-L_{\tau}}$ symmetric phase. As discussed in Ref.~\cite{Eijima:2023yiw}, this deficit can lead to the failure of the leptogenesis scenario if the breaking of $U(1)_{L_{\mu}-L_{\tau}}$ symmetry occurs subsequent to the decoupling of the sphaleron processes. This limitation arises because the production of lepton asymmetry becomes unattainable after the temperature at which the sphaleron processes decouple, denoted as $T_{\mathrm{sph}}$. The condition $T_{\mrm{bre}} < T_{\mrm{sph}}$, where $T_{\mrm{bre}} \simeq \langle \sigma \rangle$, is satisfied for $\lambda \lesssim 8 \times 10^{-7}$ for the parameter point located in the central region of the $\phi-\theta$ plane. We note that the breaking scale of the $U(1)_{L_{\mu}-L_{\tau}}$ symmetry is approximately determined by $\lambda$ in our model, as previously mentioned.

However, in the following section, we will demonstrate that the introduction of a $U(1)_{B-L}$ breaking scalar, denoted as $\Phi$, and the assumption of kinetic misalignment of $\Phi$ in the early universe can resolve this issue. This is because the kinetic motion of the scalar field can dynamically generate the CP phase, as in the Affleck-Dine mechanism~\cite{Affleck:1984fy, Dine:1995kz}. While our scenario draws inspiration from the recent work~\cite{Chun:2023eqc}, we incorporate different phenomenological aspects, as elucidated below.

\section{Leptogenesis driven by majoron rotation}
\label{sec: LG from Majoron}
In this section, we will discuss leptogenesis in the $U(1)_{L_{\mu}-L_{\tau}}$ symmetric phase, and evaluate the produced baryon asymmetry and dark matter abundance in our model. 

\subsection{Brief overview}
Firstly, we briefly explain how to realize successful leptogenesis in the $U(1)_{L_{\mu}-L_{\tau}}$ symmetric phase. 
Hereafter, we parameterize the $U(1)_{B-L}$ breaking field as 
\begin{align}
\Phi = {1 \over \sqrt{2}} \left(S + f\right) e^{i \chi/f},
\end{align}
where $S$ represents the radial mode and $\chi$ is the majoron field. (Here, $\vev{\Phi}=f/\sqrt{2}$.) The $B-L$ charge in the scalar sector, denoted as $n_{B-L}$, is given as $n_{B-L} := 2 i (\Phi \dot{\Phi}^{*}- \dot{\Phi} \Phi^{*}$), with a dot indicating a time derivative. After the radial component of the singlet scalar $\Phi$ has settled to the minimum, the $B-L$ charge asymmetry can be expressed as 
\begin{align}
n_{B-L} =  2 \dot{\Theta} f^2,
\end{align}
where $\Theta:=\chi/f$.

Suppose that the non-zero $\dot{\Theta}$ background is induced from the rotation of the complex scalar $\Phi$ in the early universe. This situation can be realized through the  kinetic misalignment mechanism~\cite{Co:2019jts,Chang:2019tvx,Co:2019wyp}\footnote{
\label{footnote: majoron oscillation}
The non-zero $\dot{\Theta}$ background could also be induced from majoron oscillation. This mechanism is called spontaneous leptogenesis~\cite{Ibe:2015nfa, Chun:2023eqc}. (See also Ref.~\cite{Kusenko:2014uta,Domcke:2020kcp}) However, to generate a sufficient amount of baryon asymmetry, a much heavier right-handed neutrino mass $M_i > 10^{10}~\mrm{GeV}$ is required~\cite{Ibe:2015nfa, Chun:2023eqc}. In contrast, in this work, we consider a lower mass scale $M_{i} \sim O(10-100)~\mrm{GeV}$.
}. As long as this rotation persists, $B-L$ charge is stored in the scalar sector, with a portion of this charge contributing to lepton asymmetry through interactions between right-handed neutrinos and leptons~\cite{Chun:2023eqc}. Importantly, even without a CP-violating decay process, lepton asymmetry can be produced. This is because the dynamical background field, the majoron $\chi$, dynamically introduces the CP phase. This becomes evident when we redefine fermionic fields $\psi \to \psi ~e^{i(B-L)_{\psi} \Theta/2}$, where $(B-L)_{\psi}$ denotes the $B-L$ number of the particle $\psi$ (See Table~\ref{tab: model}). As a consequence of this redefinition, we can eliminate the majoron field (or $\Theta$ dependence) in the original Lagrangian, and the derivative coupling with the majoron remains, i.e. $-\partial_{\mu} \Theta J^{\mu}_{B-L}/2$, where $J^{\mu}_{B-L}$ is the $B-L$ current. This derivative coupling to the majoron induces level splittings between leptons and anti-leptons~\cite{Ibe:2015nfa, Chun:2023eqc}, which leads to deviation of the thermal averaged interaction rate between the lepton number violating process and the inverse process~\cite{Ibe:2015nfa}. This deviation causes an external chemical potential~\cite{Ibe:2015nfa, Chun:2023eqc}, which can be converted to lepton asymmetry via Yukawa interaction of right-handed neutrinos with leptons and Higgs. Therefore, this mechanism offers a solution to the challenge of leptogenesis in the $U(1)_{L_{\mu}-L_{\tau}}$ symmetric phase, addressing the lack of a CP phase in the neutrino sector. The produced lepton asymmetry will be transformed into baryon asymmetry via the sphaleron process. It should be noted that this dynamic level splitting violates the CPT-invariance and hence baryon asymmetry can be generated without satisfying Sakharov’s conditions exactly~\cite{Ibe:2015nfa}.

A similar mechanism was initially proposed in the recent paper~\cite{Chun:2023eqc}, wherein the authors considered a model with purely global $U(1)_{B-L}$ symmetry and assumed the absence or substantial suppression of CP-violating decay of right-handed neutrinos. In contrast, our work explores a model that features both gauged $U(1)_{L_{\mu}-L_{\tau}}$ and global $U(1)_{B-L}$ symmetries. Consequently, i) the lack of a CP phase arises due to the restriction imposed by gauge symmetry rather than as an assumption. ii) Our model retains predictivity for neutrino oscillation parameters such as the phase of the PMNS matrix and the sum of light neutrino masses due to gauge symmetry constraints. iii) The mass spectrum of right-handed neutrinos and their interactions with standard model particles are entirely determined by three parameters: $\lambda$, $\theta$, and $\phi$. This determination remains valid even in the $U(1)_{L_{\mu}-L_{\tau}}$ symmetric phase. Moreover, we emphasize that our scenario provides a solution for realizing leptogenesis in the $U(1)_{L_{\mu}-L_{\tau}}$ symmetric phase.

\subsection{Evaluation of baryon asymmetry and dark matter abundance}
\label{subsection: evaluation}
Next, we evaluate the expected baryon asymmetry in our model. As a first step, we estimate the $B-L$ charge in the scalar sector, $n_{B-L}$, by following Refs.~\cite{Co:2019jts, Co:2019wyp, Co:2020jtv}. After that, we compute the baryon asymmetry and dark matter density produced by our model.

In this study, we employ the quartic potential previously demonstrated in Refs.~\cite{Co:2019jts, Co:2019wyp, Co:2020jtv}, given by:
\begin{align}
\label{eq: quartic potential}
V(\Phi)=\kappa^2 \left(|\Phi|^2-{f^2 \over 2}\right)^2, \quad \kappa^2 ={1\over 2} {m_{S}^2 \over f^2}.
\end{align}
Here, $m_S$ represents the vacuum mass of the radial mode of $S$. We assume that cross terms with other scalars (e.g., $ |\Phi|^2 |H|^2$, $|\Phi|^2 |\sigma_{\mu}|^2$, and $|\Phi|^2 |\sigma_{\tau}|^2$, etc.) are negligible. Additionally, we assume a large initial field value $|\Phi_i| = S_{i}/\sqrt{2} \gg f$, which is permissible when the quartic coupling is sufficiently small, indicative of a flat potential for $S$.\footnote{The flat potential naturally arises in supersymmetric theories and plays an essential role in the Affleck-Dine mechanism~\cite{Affleck:1984fy, Dine:1995kz}. In supersymmetric theories, the large field value may arise during inflation due to the negative Hubble-induced mass term~\cite{Dine:1995uk, Dine:1995kz}. In this paper, supersymmetry is not necessarily assumed. See Refs.~\cite{Co:2019jts, Co:2019wyp,Co:2020jtv} for detailed discussions and evaluations of the $U(1)$ charge in the scalar sector in supersymmetric theories.} The radial component initiates oscillations when the Hubble friction weakens compared to the curvature of the quartic potential, given by $3H(T_{S, \mrm{osc}}) \simeq \sqrt{3} \kappa S_i$. Suppose that the universe is a radiation-dominated universe at this stage. Then this condition yields $T_{S, \mrm{osc}}$ as\footnote{
The evaluation for the case that $S$ dominates the universe is given in Refs.~\cite{Co:2019jts, Co:2019wyp}, in details.
}:
\begin{align}
\label{eq: Oscillation temperatute}
T_{S,\mrm{osc}} \simeq \left({30 \over \pi^2 g_{*}}\right)^{1/4} \sqrt{\kappa M_{\mrm{pl}} S_i},
\end{align}
where $g_{*}$ represents the effective degrees of freedom, and $M_{\mrm{pl}}$ denotes the reduced Planck mass. We note that $H(T_{\mathrm{osc}})$ is constrained by the upper limit of the Hubble parameter during inflation, $H_{\mathrm{inf}} \lesssim 2.5 \times 10^{-5} M_{\mathrm{pl}}$~\cite{Planck:2018jri}. This constraint yields 
\begin{align}
\label{eq: upper limit of kappa}
\kappa \lesssim 2.5 \sqrt{3} \times 10^{-5} \left(M_{\mrm{pl}}\over S_i\right)
\end{align}
or, equivalently, $m_{S}/f \lesssim 2.5 \sqrt{6} \times 10^{-5}(M_{\mrm{pl}}/S_i)$.

To produce $B-L$ asymmetry, we introduce explicit global symmetry breaking through a higher-dimensional operator:
\begin{align}
\label{eq: higher dimensional operator}
V(\Phi)_{\cancel{B-L}}=c_n {\Phi^n \over M^{n-4}} + h.c,
\end{align}
where $M$ is a cut-off parameter and $c_n$ is a coefficient of this term. For large $S_i$, this term gives a kick to the phase direction of the scalar $\Phi$ and causes the rotation. As $S$ decreases, the explicit breaking effect diminishes, and the effect can be rephrased in terms of majoron mass, which can be estimated as follows:
\begin{align}
\label{eq: majoron mass estimation}
m_{\chi} \simeq \frac{n}{2^{(n-2)/4}} \left(|c_n| {f^{n-2} \over M^{n-4}}\right)^{1/2}.
\end{align}
Due to this potential curvature at the minimum, the $B-L$ charge might undergo a slight change after $\Phi$ reaches its minimum. However,  for $|\Phi| \sim f$, this effect is suppressed by a factor of $(f/S_{i})^n$ relative to the effect at $|\Phi| \sim S_{i}$. Therefore, in this paper, we neglect this effect and treat $n_{B-L}$ as an approximately conserved charge during leptogenesis.

If the universe remains radiation-dominated throughout its evolution, the $B-L$ charge density normalized by the entropy density, denoted as $Y_{B-L}:= n_{B-L}/s$, remains constant in the expanding universe. It is given by~\cite{Co:2019jts, Co:2019wyp, Co:2020jtv}
\begin{align}
\label{eq: B-L at the RD universe}
Y_{B-L}  = \epsilon \frac{V(\Phi_i)}{s(T_{S, \mrm{osc}}) \, m_S (\Phi_i)} = \epsilon \left( \frac{\pi^2 g_*}{30} \right)^{ \scalebox{1.01}{$\frac{3}{4}$} } \frac{ 15 \sqrt{3} }{ 4 \pi^2 g_* } \frac{ S_i^{3/2}}{\sqrt{\kappa} M_{\mrm{pl}}^{3/2} } ,
\end{align}
where $\Phi_i$ represents the initial field configuration of $\Phi$, and $\epsilon$ is defined as the ratio of the number density of phase component $n_{B-L}$/2 to the number density of the radial component $n_{S}$:
\begin{align}
\label{eq: definition of ep}
\epsilon:={n_{B-L} \over 2n_{S}},
\end{align}
with $0 < \epsilon < 1$ by definition. To evaluate $\epsilon$, we can derive the following expression for the $B-L$ asymmetry at time $t$ using the equation of motion for $\Phi$~\cite{Co:2020jtv}:
\begin{align}
\label{eq: B-L charges at t}
R^3 n_{B-L}={4n}\int R^2 \mrm{Im}\left[c_n {\Phi^n \over M^{n-4}}\right]{dR \over H},
\end{align}
where $R$ denotes the scale factor. Initially, $|\Phi|$ remains frozen due to Hubble friction, resulting in a constant field value $S = S_i$. In a radiation-dominated universe, the term on the right-hand side becomes proportional to $R^5$, making its dominant contribution evident at later times. After $3H \simeq \sqrt{3}\kappa S_i$, the time when field $S$ starts oscillating, its amplitude begins to decrease. Consequently, the contribution to the $B-L$ asymmetry diminishes at later times. This leads to the effective conservation of the number density $n_{B-L}$, as $V(\Phi)_{\cancel{B-L}}$ becomes insignificant. Thus, the $B-L$ asymmetry is predominantly generated at the initiation of $S$ oscillations at $R = R_{\mrm{osc}}$, and can be approximated by evaluating the integral at $R_{\mrm{osc}}$~\cite{Co:2020jtv}:
\begin{align}
\label{eq: B-L charges at tosc}
&n_{B-L}(t_{\rm osc}) \simeq \frac{4 \sqrt{3} n |c_n| S^{n}_{i} \delta_{\mrm{eff}}}{2^{n/2} M^{n-4} \kappa S_i},\\
\label{eq: def of deltaeff}
&\delta_{\mrm{eff}}:=\sin(n\Theta_{\rm osc}+\mrm{arg}[c_n]),
\end{align}
where $\Theta_{\rm osc}$ represents the phase of $\Phi$ when its radial component begins to oscillate, and $t_{\mrm{osc}}$ denotes the time corresponding to $R_{\mrm{osc}}$.
From this analysis, the expression for $\epsilon$ is given by
\begin{align}
\label{eq: epsilon expression}
\epsilon &\simeq {24 n \over 2^{n/2}} {|c_n| \over M^{n-4}} {S_i^{n-4} \over \kappa^2} \delta_{\mrm{eff}}\\
\label{eq: epsilon expression2}
        & \simeq   {24 \over n} {m_{\chi}^2 \over m_S^2} {S_i^{n-4} \over f^{n-4}} \delta_{\mrm{eff}},
\end{align}
where the case for $\epsilon > 1$ may seem unphysical. In reality, when the effects of higher-dimensional terms are significant, the phase degree of freedom reaches its minimum before the radial component begins to oscillate. Specifically, $\Theta_{\rm osc} \to (-\mathrm{arg}[c_n] + (2m+1)\pi)/n$ (where $m = 0, 1, 2, \dots$). As a result, the originally $O(1)$ value of $\delta_{\rm eff}$ vanishes, the rotation is suppressed, and $\epsilon \sim 0$. Therefore, in practice, $\epsilon > 1$ will not be achieved. A more detailed quantitative estimation is deferred to the subsequent discussion.

We now discuss the constraint imposed by the higher-dimensional operator introduced in Eq.~\eqref{eq: higher dimensional operator}. Generally, such an operator can potentially induce an unwanted minimum in the potential of $\Phi$ and may spoil this mechanism. This risk arises because the scalar field $\Phi$ possesses a large initial field value and might eventually settle into this minimum~\cite{Kawasaki:2006yb, Enomoto:2023sma}. To circumvent this issue, at the onset of oscillations, the curvature in the phase direction needs to be smaller than the Hubble parameter. This condition is given by the following expression~\cite{Enomoto:2023sma}:
\begin{align}
\label{eq: early trapping condition}
m_{\chi,\rm{eff}}(S_i) \lesssim H(T_{\rm{S,osc}}),
\end{align}
where $m_{\chi,\rm{eff}}(S_i):=m_{\chi}(S_{i}/f)^{(n-2)/2}$. This condition can also be interpreted as the requirement, that the mass of the radial mode evaluated at the large initial field value $S_i$ should be larger than the mass of the angular mode evaluated at $S_i$, which is the condition for the validity of the Nambu-Goldstone treatment~\cite{Berbig:2023uzs}.

For the quartic potential given by Eq.~\eqref{eq: quartic potential}, this condition simplifies to the upper limit on the majoron mass,
\begin{align}
\label{eq: upper limit of majoron mass}
m_{\chi} \lesssim 10^{23-5n}~\mrm{GeV}\left(\kappa \over 10^{-5}\right)\left(S_{i}\over M_{\mrm{pl}}\right)^{-\frac{n-2}{2}}\left(f\over10^{8}~\mrm{GeV}\right)^{\frac{n-2}{2}}\left(10^{18}~\mrm{GeV}\over M_{\mrm{pl}}\right)^{\frac{n-4}{2}}.
\end{align}
For the cases $n=5$ and $n=6$, this yields $m_{\chi} \lesssim 10~\mathrm{MeV}$ and $m_{\chi} \lesssim 100~\mathrm{eV}$ respectively, for $f \simeq 10^{8}~\mathrm{GeV}$ and $\kappa \simeq 10^{-5}$. This constraint, Eq.~\eqref{eq: early trapping condition}, also establishes an upper limit on $\epsilon$ as 
\begin{align}
\label{eq: upper limit of epsilon}
\epsilon \lesssim \frac{4}{n}\delta_{\mrm{eff}},
\end{align}
from Eq.~\eqref{eq: epsilon expression2}. Therefore, $\epsilon$ cannot be arbitrarily large. In general, the larger the value of $n$, the more stringent the constraints on $\epsilon$ become. If parameters that do not satisfy this constraint are chosen, the phase direction will reach its minimum before the radial direction, as mentioned above, resulting in $\delta_{\mrm{eff}} \ll 1$ and hindering the rotation.

If the right-handed neutrinos are in the thermal bath, the non-zero $\dot{\Theta}$ background or equivalently, the produced $B-L$ asymmetry in the scalar sector, causes lepton asymmetry in the SM sector through the Yukawa interaction of the right-handed neutrino. To analyze this, we diagonalize the Majorana mass term in the $U(1)_{L_{\mu}-L_{\tau}}$ symmetric phase~\eqref{eq: Majorana mass matrix in symmetric phase} as 
\begin{align}
\mathcal{L}_N  = 
&-\hat{\lambda}_{i\alpha} \hat{N}_{i}^{c} (L_{\alpha} \cdot H) -\frac{1}{2} M_{i}^{\mrm{sym}} \hat{N}_{i}^{c} \hat{N}_{i}^{c} + \text{h.c.}~,
\label{eq: Lagrangian_MR_diag}
\end{align}
where,
\begin{align}
{\cal M}_{R, \mrm{sym}} &= \Omega_{\mrm{sym}}^{\ast} \text{diag}(M_{1}^{\mrm{sym}},M_{2}^{\mrm{sym}},M_{3}^{\mrm{sym}}) \Omega_{\mrm{sym}}^{\dagger}~, \\
\hat{N}_{i}^{c} &= \sum_{\alpha} \Omega_{\mrm{sym},\alpha i}^{\ast} N_{\alpha}^{c}~, \\
\hat{\lambda}_{i\alpha} &= \Omega_{\mrm{sym},\alpha i} \lambda_{\alpha}~(\text{not~summed})~,
\end{align}
where $\Omega_{\mrm{sym}}$ is a unitary matrix that diagonalizes ${\cal M}_{R, \mrm{sym}}$, and $M_{1,\,2,\,3}^{\mrm{sym}}$ are the right-handed neutrino masses in the $U(1)_{L_{\mu}-L_{\tau}}$ symmetric phase. 

\begin{figure}[t!]
    \centering
    \includegraphics[width = .48\textwidth]{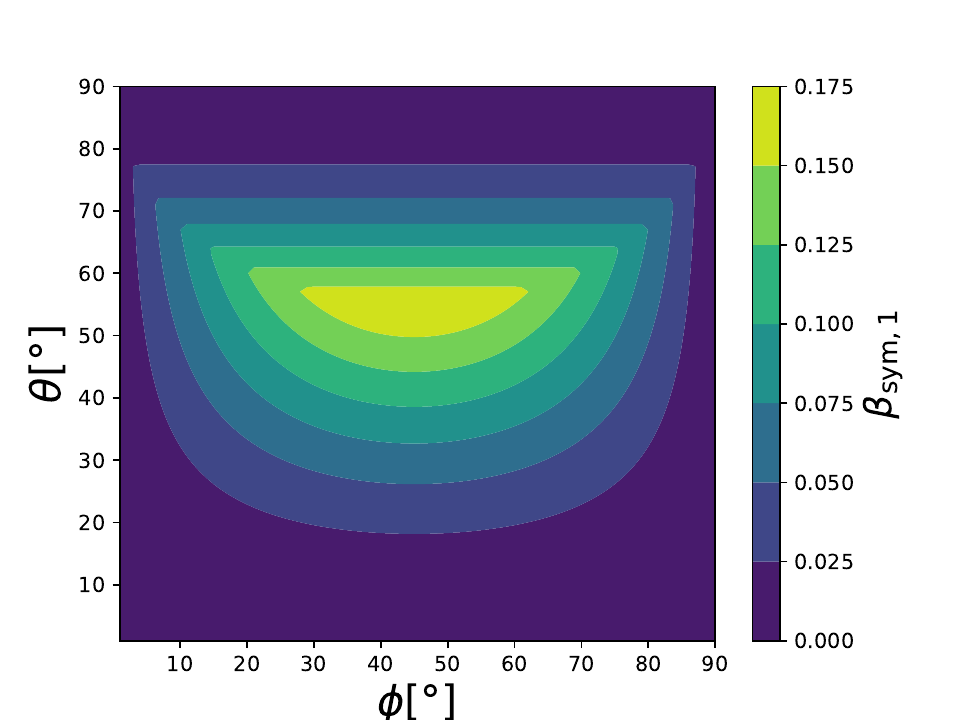}
    
    \includegraphics[width = .48\textwidth]{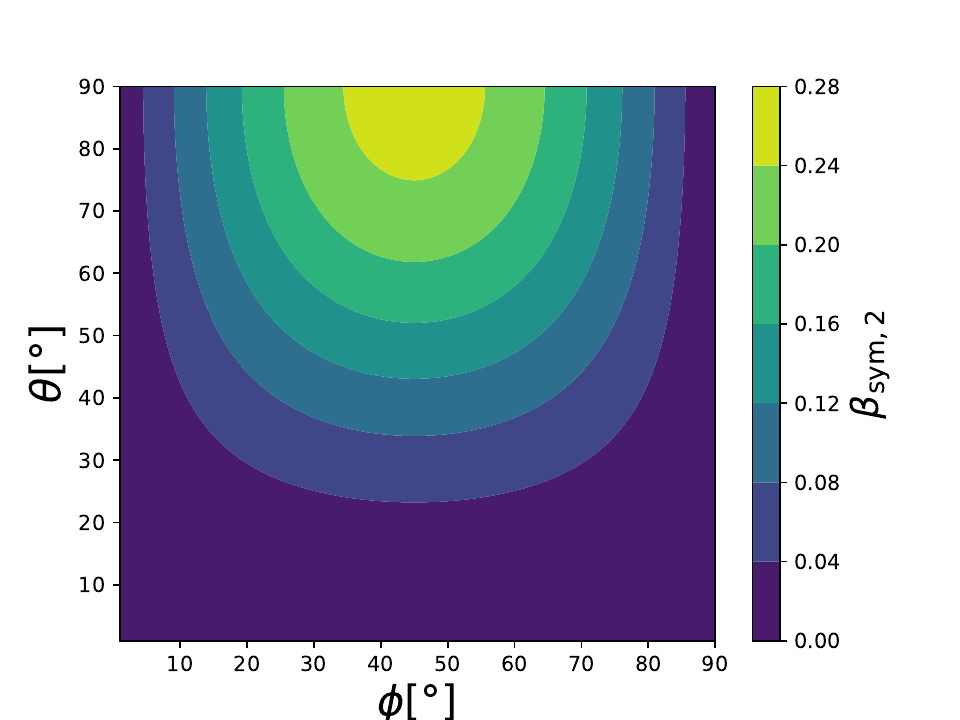}
    \includegraphics[width = .48\textwidth]{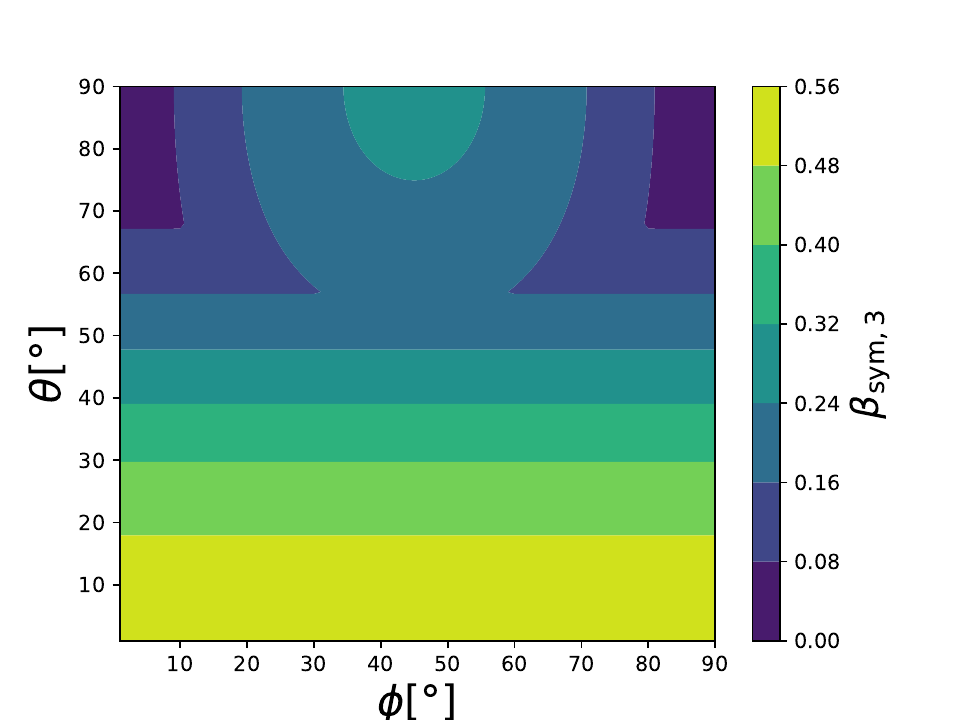}
    \caption{Contour plots illustrating the behavior of the functions $\beta_{\mrm{sym},1}$ (top panel), $\beta_{\mrm{sym},2}$ (bottom left), and $\beta_{\mrm{sym},3}$ (bottom right). The input parameter values are consistent with those used in Fig.~\ref{fig: Majorana mass scan}. It is noteworthy that for small values of $\theta$, corresponding to a large $M_{ee}$, the mass of the heaviest right-handed neutrino $M_3$ can be approximated as $M_{ee} \propto \cos^2 \theta$ and remains independent of $\phi$.}
    \label{fig: Contour_beta}
\end{figure}

It is worth noting that the components of the Majorana mass ${\cal M}_{R, \mrm{sym}}$, are equivalent to those of ${\cal M}_{R}$ as defined in Eq.~\eqref{eq: Dirac and Majorana mass}. Consequently, these components can be computed from the light neutrino mass matrix using Eq.~\eqref{eq: seesaw inverse}. As a consequence, the masses of the right-handed neutrinos are proportional to $\lambda^2$ even in the symmetric phase. This dependence can be expressed more precisely as follows,
\begin{align}
\label{eq: RHN masses}
M_{1,\,2,\,3}^{\mrm{sym}} = \frac{v^2 \lambda^2}{2m_1}\beta_{\mrm{sym},1,\,2,\, 3} , 
\end{align}
where $\beta_{\mrm{sym},1,\,2,\, 3}$ are real numbers satisfying $\beta_{\mrm{sym},1,\,2,\, 3} \lesssim O(1)$. We present contour plots of $\beta_{\mrm{sym},1,\,2,\, 3}$ in the $\phi-\theta$ plane in Fig.~\ref{fig: Contour_beta}. These plots utilize the same neutrino oscillation parameters as in Fig.~\ref{fig: Majorana mass scan}. As indicated in Eq.~\eqref{eq: Majorana mass matrix in symmetric phase}, the two mass eigenvalues are exactly degenerate, i.e., $M_1^\mrm{sym} < M_2^\mrm{sym} =M_3^\mrm{sym}$ or $M_1^\mrm{sym} = M_2^\mrm{sym} < M_3^\mrm{sym}$, depending on the values of $\theta$ and $\phi$. 

The decay width of the right-handed neutrino to the lepton and SM Higgs boson in $U(1)_{L_{\mu}-L_{\tau}}$ symmetric phase is given by
\begin{align}
\Gamma^{\mrm{sym}}_{N_i \to l_{\alpha} H} = {1 \over 8 \pi} (\hat{\lambda}_{i\alpha}\hat{\lambda}^*_{\alpha i})M_i^{\mrm{sym}},
\end{align}
and total decay width to the SM leptons is $\Gamma^{\mrm{sym}}_{N_i \to l H}:=\sum_{\alpha} \Gamma^{\mrm{sym}}_{N_i \to l_{\alpha} H}={1 \over 8 \pi} (\hat{\lambda}\hat{\lambda}^{\dagger})_{ii}M_i^{\mrm{sym}}$. It is convenient to define the thermal averaged decay width as follows:
\begin{align}
\label{eq: thermal averaged interaction rate}
\vev{\Gamma^{\mrm{sym}}_{N_i \to l_{\alpha} H}}:= {K_1(z)\over K_2(z)}\Gamma^{\mrm{sym}}_{N_i \to l_{\alpha} H},
\end{align}
with $K_{1}$ and $K_{2}$ are modified Bessel functions for 1st and 2nd kind, and $z:=M_1^{\mrm{sym}}/T$. 

In the presence of a non-zero $\dot{\Theta}$ background, the Boltzmann equation for the lepton asymmetry density $n_{\Delta l}:= n_{l}-n_{\bar{l}}$ is modified as follows~\cite{Domcke:2020quw, Chun:2023eqc}\footnote{Since we focus on low-scale leptogenesis $M_i^{\mrm{sym}} \ll 10^5~\mathrm{GeV}$, we neglect the heavy flavor effect~\cite{Barbieri:1999ma, Engelhard:2006yg, Antusch:2010ms, Blanchet:2011xq} when evaluating
baryon (lepton) asymmetry, while flavor effect is taken into account~\cite{Barbieri:1999ma, Abada:2006fw, Nardi:2006fx}. In our model, another type of flavor effect potentially exists, caused by $U(1)_{L_{\mu}-L_{\tau}}$ gauge bosons and Higgs bosons because they are also in the thermal bath. However, a detailed analysis of this effect goes beyond the scope of our paper.}:
\begin{align}
\label{eq: Boltzmann equation}
\dot{n}_{\Delta l_{\alpha}} + 3 H n_{\Delta l_{\alpha}} = -\sum_{i} n_{N_i}^{\mrm{eq}}\vev{\Gamma^{\mrm{sym}}_{N_i \to l_{\alpha} H}} \left(\frac{n_{\Delta l_\alpha}}{n_{l_\alpha}^{\mrm{eq}}} +\frac{n_{\Delta H}}{n_H^{\mrm{eq}}} -\frac{\dot{\Theta}}{T}\right), ~(\alpha = e,\mu, \tau)
\end{align}
where $n_{\Delta H} := n_{H} - n_{\bar{H}}$, and $n_{X}, (X=l_{\alpha},N, H)$ represent the equilibrium number density of $X$. Here, we neglect the lepton number violating scattering term because they are a higher order of $\lambda \ll 1$. We briefly show the derivation of this Boltzmann equation in Appendix~\ref{appendix: derivation of Boltzmann equation}. 

Significantly, in our model during leptogenesis, the right-handed neutrinos are in thermal equilibrium, i.e., $n_{N_i}=n_{N_i}^{\mathrm{eq}}$. Additionally, the absence of CP phases in the neutrino sector results in the lack of a decay term. Consequently, only terms corresponding to the inverse decay remain on the right-hand side of Eq.~\eqref{eq: Boltzmann equation}.
As shown in Eq.~\eqref{eq: Boltzmann equation}, if inverse decay is in thermal equilibrium, the lepton asymmetry becomes:
\begin{align}
n_{\Delta l} = \frac{c_{L}}{6} \dot{\Theta} T^2,
\end{align}
where $c_{L}$ is the coefficient determined by the equations of chemical equilibrium. At low temperatures, $T < 10^5~\mathrm{GeV}$, $c_L \simeq 51/22$~\cite{Chun:2023eqc}. In comparison to the asymmetry in the scalar sector $n_{B-L} = 2\dot{\Theta} f^2$, the lepton asymmetry in the thermal bath is suppressed by the factor $T^2/f^2$. 

Since the production of lepton asymmetry in our model relies on the inverse decay of right-handed neutrinos, a strong washout condition is required for this mechanism. This condition is quantified by the decay parameter $K:=\tilde{m}_1/m_{*} >1$~\cite{Buchmuller:2004nz}, where $\tilde{m}_1:=(\hat{\lambda}\hat{\lambda}^{\dagger})v^2/2M_1$, and $m_{*}\simeq 1\times 10^{-3}~\mathrm{eV}$ represent the effective neutrino mass, and the equilibrium neutrino mass respectively. In our model, the lightest neutrino mass is on the order of the atmospheric neutrino mass, i.e., $m_1 \simeq O(0.05)~\mathrm{eV}$, indicating that the effective mass $\tilde{m}_1$ is typically of the same order. Consequently, the strong washout condition is indeed satisfied with $K > 1$. To assess the baryon asymmetry, it is crucial to identify the temperature range where the inverse decay process of the right-handed neutrino is in thermal equilibrium. To do this, it is convenient to define the following conventional function,
\begin{align}
W_{\mrm{ID}}(z)
:=\frac{1}{2} \sum_{i}\frac{\vev{\Gamma^{\mrm{sym}}_{N_i \to l H}}}{H(z)z}\frac{n^{\mrm{eq}}_{N_i}}{n^{\mrm{eq}}_{l_{\alpha}}},
\end{align}
which parametrizes whether inverse right-handed neutrino decay is in thermal equilibrium or not, that is, if $W_{\mrm{ID}}(z) > 1$, the inverse right-handed neutrino decay is in thermal equilibrium. In our scenario, we focus on the centered region in the $\theta-\phi$ plane, where the mass eigenvalue of each right-handed neutrino is in the same scale, $M_1\simeq M_2 \simeq M_3$, and also, the decay rate is approximately same as well. Therefore, $W_{\mrm{ID}}(z)$ can be approximately evaluated by
\begin{align}
W_{\mrm{ID}}(z)
&\simeq \frac{3}{2} \frac{\vev{\Gamma^{\mrm{sym}}_{N_1 \to l H}}}{H(z)z}\frac{n^{\mrm{eq}}_{N_1}}{n^{\mrm{eq}}_{l_{\alpha}}}\nn
&=\frac{3}{4} K z^3 K_1(z).
\end{align}
Given a fixed value of $K$, the condition $W_{\mrm{ID}}(z) > 1$ leads to the range:
\begin{align}
M_1^{\mrm{sym}}/z_{\mrm{out}} \lesssim T \lesssim M_1^{\mrm{sym}}/z_{\mrm{in}},
\end{align}
where $z_{\text{in}}$ and $z_{\text{out}}$ represent the minimum and maximum values of $z$ at which the inverse right-handed neutrino decay is in thermal equilibrium, respectively. Therefore, leptogenesis occurs in the interval $z_{\text{in}} < z < z_{\text{out}}$ within our model. We have numerically computed these values for our benchmark points and found that $M_1^{\text{sym}}/z_{\text{in}} > T_{\text{sph}}$ is satisfied for these points, as we will explain in the subsequent section.

Finally, the lepton asymmetry in the thermal bath will be converted to the baryon asymmetry via the sphaleron process. 
Assuming the right-handed neutrino inverse decay is in thermal equilibrium during leptogenesis, the resulting baryon asymmetry $Y_{B}:=n_{B}/s$ is fixed at the temperature where the sphaleron process freezes out. For the case $n=5$, the evaluation of $Y_{B}$ is given by
\begin{align}
\label{eq: BAU n=5}
Y_{B}&={1\over6}c_{B} Y_{B-L} \left({T_{\mrm{sph}} \over f}\right)^2 \nn
     &\simeq 9 \times 10^{-11} \left({\delta_{\mrm{eff}} \over 1}\right)
     \left({T_{\mrm{sph}} \over130~\mrm{GeV}}\right)^{2}\left({S_i \over M_{\mrm{pl}}}\right)^{5/2} \nn
     &\quad \times
     \left({M_{\mrm{pl}} \over 2.4 \times 10^{18}~\mrm{GeV}}\right)
     \left({10^{-2}~\mrm{GeV} \over m_S}\right)^{5/2}
     \left({m_{\chi}\over 0.5\mrm{eV}}\right)^{2}
     \left({10^7~\mrm{GeV} \over f}\right)^{5/2},
\end{align}
and for the case $n=6$,
\begin{align}
\label{eq: BAU n=6}
Y_{B}&={1\over6}c_{B} Y_{B-L} \left({T_{\mrm{sph}} \over f}\right)^2 \nn
     &\simeq 9 \times 10^{-11} \left({\delta_{\mrm{eff}} \over 1}\right)
     \left({T_{\mrm{sph}} \over130~\mrm{GeV}}\right)^{2}\left({S_i \over M_{\mrm{pl}}}\right)^{7/2} \nn
     &\quad \times
     \left({M_{\mrm{pl}} \over 2.4 \times 10^{18}~\mrm{GeV}}\right)^{2} 
     \left({2 \times 10^{-2}~\mrm{GeV} \over m_S}\right)^{5/2}
     \left({m_{\chi}\over 0.05\mrm{meV}}\right)^{2}
     \left({3 \times 10^8~\mrm{GeV} \over f}\right)^{7/2},
\end{align}
where $c_{B}$ is a temperature-dependent coefficient, $Y_{B-L}$ is given by Eq~\eqref{eq: B-L at the RD universe}, and $\delta_{\mrm{eff}}$ is defined in Eq.~\eqref{eq: def of deltaeff}. We have used the value $c_{B}$ at low temperatures, specifically $T < 10^5~\mathrm{GeV}$, which is $c_{B}=-(28/22)$~\cite{Chun:2023eqc}. We have also used Eq.~\eqref{eq: epsilon expression} to evaluate the $\epsilon$. In the next section, we will present the parameter region capable of producing the correct baryon asymmetry in Fig~\ref{fig: BAU}. If the right-handed neutrino is too light such that $M_1^{\mrm{sym}}/z_{\mrm{in}} < T_{\mrm{sph}}$, the inverse decay is not in thermal equilibrium at the sphaleron decoupling temperature, hence baryon asymmetry is produced through the freeze-in mechanism. In this case, we have an additional suppression factor, expected to be $\sim W_{\mrm{ID}}(z_{\mrm{sph}})z_{\mrm{sph}}$, where $z_{\mrm{sph}}=M_{1}^{\mrm{sym}}/T_{\mrm{sph}}$~\cite{Chun:2023eqc, Barnes:2024jap}. We need to care about this effect when $M_{1}^{\mrm{sym}} \lesssim 20~\mrm{GeV}$.

Next, we evaluate the dark matter abundance within the context of our model. In our model, the abundance of dark matter is generated through the coherent oscillation of the majoron.\footnote{The potential overabundance due to the thermal production of majorons via $HL \leftrightarrow \chi N$ in thermal equilibrium, as discussed in~\cite{Chun:2023eqc}, is negligible in our model. This is attributed to the sufficiently suppressed coupling constants between the majoron and right-handed neutrino, as outlined in Eq.~\eqref{eq: coupling hierarchy}.} In contrast to the conventional misalignment mechanism~\cite{Preskill:1982cy, Abbott:1982af, Dine:1982ah}, the majoron field has the initial energy of kinetic motion in our model. This delays the epoch of majoron oscillation until the kinetic energy becomes comparable to the potential energy of the majoron. This mechanism is termed the kinetic misalignment mechanism~\cite{Co:2019jts, Chang:2019tvx, Co:2019wyp}. 

After sphaleron freezes out, the kinetic energy of the majoron field decreases as $f^2 \dot{\Theta^2}/2 \propto T^6$ because $n_{B-L} = 2 f^2 \dot{\Theta} \propto T^3$. Eventually, the majoron field fails to overcome the potential barrier induced by the higher-dimensional operator expressed in Eq.~\eqref{eq: higher dimensional operator} on the bottom, resulting in the cessation of rotation. We denote this temperature as $T_{\mrm{trap}}$, determined by $f^2 \dot{\Theta}^2(T_{\mrm{trap}})/2 \simeq m_{\chi}^2 f^2$. In the conventional misalignment mechanism, the majoron field starts to oscillate coherently when $3 H(T_{\chi, \mrm{osc}}) \simeq m_{\chi}$.  However, if $T_{\mrm{trap}} < T_{\chi, \mrm{osc}}$, the oscillation starts at the $T_{\mrm{trap}}$ rather than $T_{\chi, \mrm{osc}}$. Notably, $T_{\mrm{trap}}$ depends on the generated baryon asymmetry:
\begin{align}
s(T_{\mrm{trap}}) \simeq \frac{ 2f^2 m_{\chi}}{Y_{B-L}} ={c_{B} \over 3}  \frac{ T_{\mrm{sph}}^2 m_{\chi}}{Y_{B}},
\end{align}
where we note that $\dot{\Theta}(T_{\mathrm{trap}}) \simeq m_{\chi}$ holds. For the scenario where $M_1^{\mrm{sym}} /z_{\mrm{in}} > T_{\mrm{sph}}$, this results in
\begin{align}
T_{\mrm{trap}}&= \left({c_{B} \over 3} Y_{B}^{-1} \left({2 \pi^2\over 45} g_{*}\right)^{-1} m_{\chi} T_{\mrm{sph}}^2\right)^{1/3}\nn
&\simeq 10^2~\mrm{GeV}\left({m_{\chi}\over \mrm{keV}}\right)^{1/3},
\end{align}
where on the second line we have taken $Y_B = Y_B^{\mrm{osb}}\simeq 8.7 \times 10^{-11}$~\cite{Planck:2018vyg}. For successful leptogenesis, it is imperative that $T_{\mrm{trap}} < T_{\mrm{sph}}$. Otherwise, the majoron rotation stops, and we lose the source of the CP phase during the leptogenesis. This means that we need the sufficiently light majoron, $m_{\chi} \leq \mrm{keV}$. On the other hand, $T_{\chi, \mrm{osc}}$ is given by
\begin{align}
T_{\chi, \mrm{osc}} = 5 \times 10^5~\mrm{GeV}\left({m_{\chi}\over \mrm{keV}}\right)^{1/2},
\end{align}
which is much higher than $T_{\mrm{trap}}$ at the mass range of $m_{\chi}$ that we are interested in.
Therefore dark matter density in our model is evaluated in the manner of kinetic misalignment mechanism, in which we have less redshift until now:
\begin{align}
{\rho_{\mrm{\chi}}\over s} \simeq {m_{\chi}^2 f^2 \over s(T_{\mrm{trap}})},
\end{align}
where $\rho_{\mrm{\chi}}$ is energy density of majoron oscillation.
In the case of $M_1^{\mrm{sym}} /z_{\mrm{in}} > T_{\mrm{sph}}$, the majoron abundance becomes 
\begin{align}
\label{eq: dm benchmark}
{\Omega_{\chi} h^2 \over \Omega_{\mrm{DM}} h^2} \simeq \left({m_{\chi}\over 0.05~\mrm{meV}}\right)\left({f\over 6 \times 10^8~\mrm{GeV}}\right)^2,
\end{align}
with $\Omega_{\mrm{DM}} h^2 \simeq 0.12$ is observed dark matter abundance~\cite{Planck:2018vyg}.\footnote{In the case of $M_1^{\mrm{sym}} /z_{\mrm{in}} < T_{\mrm{sph}}$, we need large $\dot{\Theta}$ to generate the correct baryon asymmetry. This leads to the lower $T_{\mrm{trap}}$ and larger majoron abundance. }

In our model, the produced majoron can only decay to the neutrino pairs. The decay width of this majoron dark matter $\Gamma_{\chi}$ is given by
\begin{align}
\Gamma_{\chi} \simeq {1\over16 \pi}{m_{\chi} \over f^2} \sum_{i=1,2,3}m_i^2.
\end{align}
We note that the sum of neutrino mass squared can be predicted in our model, given as $\sum_{i}m_i^2 \simeq 0.005~\mrm{eV}^2$ from neutrino oscillation data in the \texttt{NuFit 5.2} analysis~\cite{nufit, Esteban:2020cvm} (without SK data). The lifetime of this majoron dark matter, denoted as $\tau_{\chi}$ and equal to $\Gamma_{\chi}^{-1}$, is tightly restricted from CMB and BAO analysis~\cite{Audren:2014bca,Enqvist:2019tsa,Nygaard:2020sow,Alvi:2022aam,Simon:2022ftd}, and the lower limit is $\tau_{\chi} > 250~\mrm{Gyr}$. This yields the following constraint to account for the dark matter abundance:
\begin{align}
\label{eq: Constraints on majoron DM}
f \gtrsim 2 \times 10^6~\mrm{GeV},~m_{\chi} \lesssim 4~\mrm{eV}.
\end{align}

Finally, By substituting Eq.~\eqref{eq: dm benchmark} into Eqs.~\eqref{eq: BAU n=5} and~\eqref{eq: BAU n=6}, we can show the parameter space where both the correct baryon asymmetry and dark matter density are achieved. For the $n=5$, the expression for the baryon asymmetry is given by
\begin{align}
\label{eq: BAU and DM n=5}
Y_{B}&={1\over6}c_{B} Y_{B-L} \left({T_{\mrm{sph}} \over f}\right)^2 \nn
     &\simeq 9 \times 10^{-11} \left({\delta_{\mrm{eff}} \over 1}\right)
     \left({T_{\mrm{sph}} \over130~\mrm{GeV}}\right)^{6} 
     \left({S_i \over M_{\mrm{pl}}}\right)^{7/2}\nn
     &\quad \times
     \left({M_{\mrm{pl}} \over 2.4 \times 10^{18}~\rm{GeV}}\right)
     \left({10^{-2}~\mrm{GeV} \over m_S}\right)^{5/2}
     \left({10^7~\mrm{GeV} \over f}\right)^{13/2}
     ,
\end{align}
and for the $n=6$,
\begin{align}
\label{eq: BAU and DM n=6}
Y_{B}&={1\over6}c_{B} Y_{B-L} \left({T_{\mrm{sph}} \over f}\right)^2 \nn
     &\simeq 9 \times 10^{-11} \left({\delta_{\mrm{eff}} \over 1}\right)
     \left({T_{\mrm{sph}} \over130~\mrm{GeV}}\right)^{6} 
     \left({S_i \over M_{\mrm{pl}}}\right)^{7/2}\nn
     &\quad \times
     \left({M_{\mrm{pl}} \over 2.4 \times 10^{18}~\rm{GeV}}\right)^{2}
     \left({2 \times 10^{-2}~\mrm{GeV} \over m_S}\right)^{5/2}
     \left({3 \times 10^8~\mrm{GeV} \over f}\right)^{15/2}
     ,
\end{align}
where the parameters are fixed to satisfy Eq.~\eqref{eq: dm benchmark}. In each case, the expression for $\epsilon$ can be obtained by substituting $m_{\chi}$, which can account for dark matter, and $m_{S}$, which can explain the BAU, into Eq.~\eqref{eq: epsilon expression2}:
\begin{align}
\epsilon
    \simeq
        \begin{cases}
		 4 \times 10^{-5} ~\delta_{\mrm{eff}}^{1/5} ~(f/\mathrm{GeV})^{1/5}, & n=5 \\
		0.2 ~\delta_{\mrm{eff}}^{1/5}, & n=6
		\end{cases}
\end{align}
thereby fulfilling the upper limit of $\epsilon$, as given in Eq.~\eqref{eq: upper limit of epsilon}.

\section{Results}
\label{sec: result}
In this section, we discuss the prediction and allowed region in our model.  Our analysis utilizes neutrino oscillation data from the \texttt{NuFit 5.2} analysis~\cite{nufit, Esteban:2020cvm} (without SK data) as input parameters. To avoid constraints on the sum of light neutrino masses, $\sum_{i} m_{i} < (0.12-0.69)~\mathrm{eV}$~\cite{LVinZyla:2020zbs, Capozzi_2020}, we set $\theta_{23}$ to the $+3\sigma$ range, consistent with Ref.~\cite{Granelli:2023egb}. This choice predicts a sum of the light neutrino masses $\sum_i m_i = 0.117~\mrm{eV}$ and the PMNS phases $\delta \simeq 228^\circ$, $\alpha_2 = 225^\circ$, and $\alpha_3 = 70^\circ$~\cite{Granelli:2023egb}. In addition to these, we introduce three free input parameters: $\lambda$, $\theta$, and $\phi$. In essence, $\lambda$ determines the typical mass scale of right-handed neutrino masses, as expressed in Eq.~\eqref{eq: RHN masses}. We concentrate on scenarios with a low $U(1)_{L_{\mu}-L_{\tau}}$ breaking scale, where $U(1)_{L_{\mu}-L_{\tau}}$ is restored in the early universe. This occurs when $\lambda \lesssim 8\times 10^{-7}$, particularly within the central region of the $\phi-\theta$ plane. To account for the muon $g-2$ anomaly, we require $\langle \sigma \rangle = 10- 50~\mathrm{GeV}$, necessitating a lower scale for $\lambda$, $\lambda \lesssim 5 \times 10^{-7}$ within the central region of the $\phi-\theta$ plane.

We examine two benchmark points: i) $\lambda=8 \times 10^{-7}$, $\theta=56^\circ$, and $\phi=38^\circ$, resulting in $\lambda_{e\mu}\langle \sigma_{\mu} \rangle \simeq \lambda_{e\tau}\langle \sigma_{\tau} \rangle \simeq 123~\mathrm{GeV}$, and the masses of right-handed neutrinos given by $M_1^\mathrm{sym} \simeq 111~\mathrm{GeV}$, $M_2^\mathrm{sym} = M_3^\mathrm{sym} \simeq 115~\mathrm{GeV}$; ii) $\lambda=4 \times 10^{-7}$, $\theta=56^\circ$, and $\phi=38^\circ$, yielding $\lambda_{e\mu}\langle \sigma_{\mu} \rangle \simeq \lambda_{e\tau}\langle \sigma_{\tau} \rangle \simeq 30~\mathrm{GeV}$, aligning with the $U(1)_{L_{\mu}-L_{\tau}}$ breaking scale preferred by the muon $g-2$ anomaly. At this point, we have $M_1^\mathrm{sym} \simeq 28~\mathrm{GeV}$, and $M_2^\mathrm{sym} = M_3^\mathrm{sym} \simeq 29~\mathrm{GeV}$. In both cases, the $U(1)_{L_{\mu}-L_{\tau}}$ breaking scale is lower than the temperature where the sphaleron process decouples ($\langle \sigma \rangle < T_{\mathrm{sph}}$), indicating leptogenesis occurs in the $U(1)_{L_{\mu}-L_{\tau}}$ symmetric phase. Also, we have verified that $M_1^{\mathrm{sym}}/z_{\mathrm{in}} > T_{\mathrm{sph}}$ holds true for both benchmark points. Consequently, we do not encounter an additional suppression factor for the produced baryon asymmetry arising from the freeze-in mechanism.

\begin{figure}[t!]
    \centering
    \includegraphics[width = .8\textwidth]{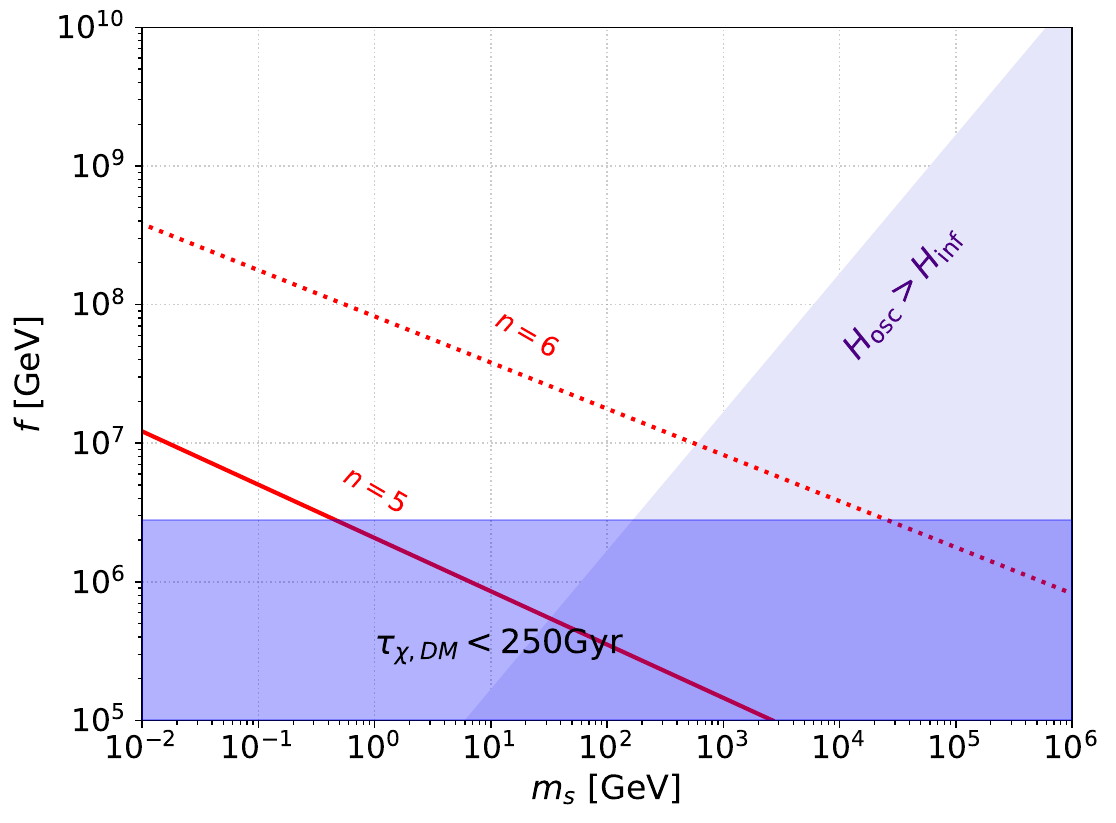}
    \caption{The parameter space where both the correct baryon asymmetry and dark matter abundance are achieved is depicted. The red and red dotted line represent the regions where the correct baryon asymmetry is obtained for $n=5$ and $n=6$ with $\delta_{\text{eff}}=1$. In the light purple region, the value of $H_{\text{osc}}:=H(T_{\text{osc}})$ surpasses $H_{\text{inf}}$ thereby not justifying the large initial field value of $\Phi$. The blue region is excluded based on constraints from CMB and BAO analyses~\cite{Audren:2014bca, Enqvist:2019tsa, Nygaard:2020sow, Alvi:2022aam, Simon:2022ftd}, which limit the lifetime of the majoron dark matter to $\tau_{\chi} < 250~\text{Gyr}$. 
}
    \label{fig: BAU}
\end{figure}

\begin{figure}[t!]
    \centering
    \includegraphics[width = .8\textwidth]{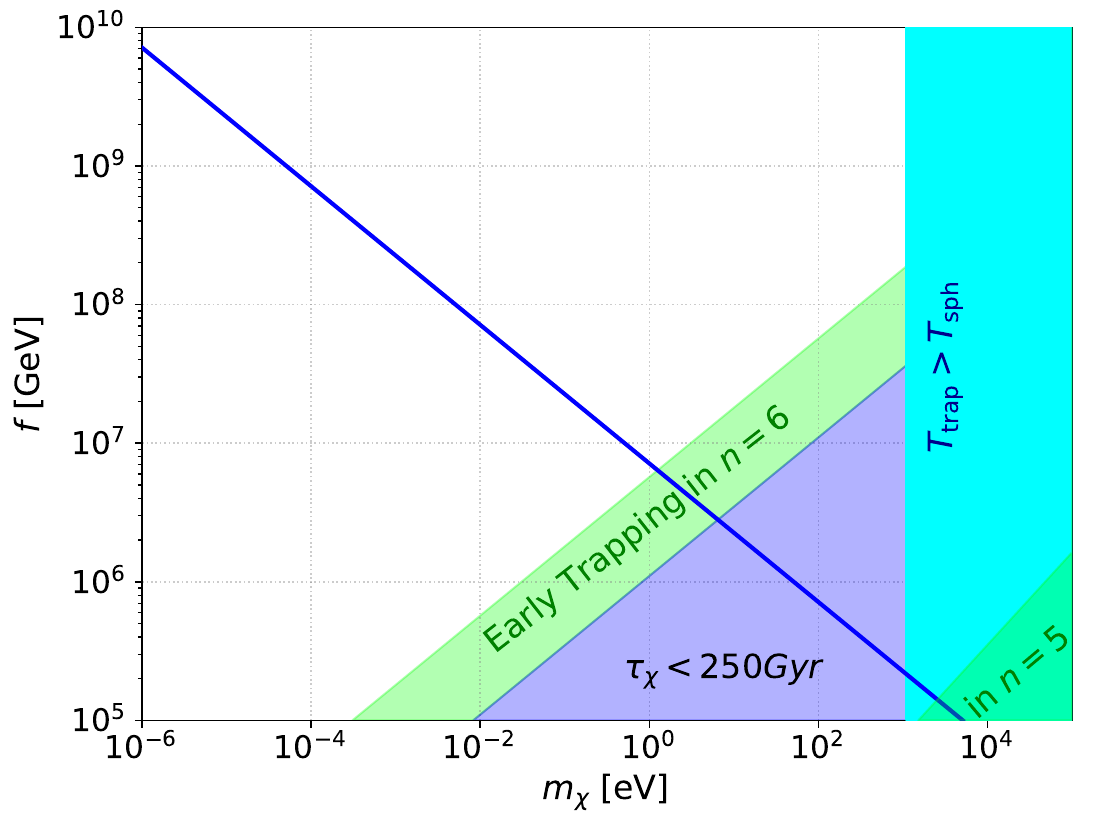}
    \caption{The parameter space that yields the correct dark matter abundance is depicted. The blue line represents the region where the correct dark matter density is achieved. Within the cyan region, the rotation of the majoron halts during leptogenesis, rendering our estimation unreliable. The green region suggests that the $U(1)_{B-L}$ breaking field might become trapped in an undesirable minimum, inhibiting its rotation. This behavior is influenced by the magnitude of the effect of the higher-dimensional operator, which gives the majoron mass. Here, we focus on $n=5$ and $n=6$ and show the case of $\kappa \simeq 10^{-5}$ as suggested by the upper limit of $\kappa$ in Eq. \eqref{eq: upper limit of kappa}. The blue region is excluded based on constraints from CMB and BAO analyses~\cite{Audren:2014bca, Enqvist:2019tsa, Nygaard:2020sow, Alvi:2022aam, Simon:2022ftd}, which limit the lifetime of the majoron dark matter to $\tau_{\chi} < 250~\text{Gyr}$.
}
    \label{fig: DM}
\end{figure}

We present our results in Figs.~\ref{fig: BAU} and~\ref{fig: DM}. In Fig.~\ref{fig: BAU}, the red and red dotted lines represent the parameter space where both the correct baryon asymmetry and dark matter density are achieved for $n=5$ and $n=6$, respectively, with $\delta_{\mrm{eff}}=1$ in our model. In the light purple region, $H_{\mrm{osc}} := H(T_{\mrm{osc}})$ surpasses $H_{\mrm{inf}}$, undermining the justification for the large initial field value of $\Phi$. The blue region is excluded based on the aforementioned CMB and BAO analyses~\cite{Audren:2014bca, Enqvist:2019tsa, Nygaard:2020sow, Alvi:2022aam, Simon:2022ftd}, which set constraints on the lifetime of the majoron dark matter: $\tau_{\chi} < 250~\text{Gyr}$.

In Fig.~\ref{fig: DM}, the blue line depicts the region where the correct dark matter abundance is achieved. Within the cyan region, the rotation of the majoron halts during leptogenesis, rendering our estimation unreliable. The blue region is again excluded based on the aforementioned CMB and BAO analyses~\cite{Audren:2014bca, Enqvist:2019tsa, Nygaard:2020sow, Alvi:2022aam, Simon:2022ftd}, which constrain the majoron dark matter lifetime to $\tau_{\chi} < 250~\text{Gyr}$. The green region indicates that the $U(1)_{B-L}$ breaking field might be trapped in an undesirable minimum, preventing its rotation. This dependency is influenced by the magnitude of the effect of the higher-dimensional operator. For instance, we illustrate the case for $n=5$ and $n=6$ in Equation~\eqref{eq: higher dimensional operator}, assuming $\kappa \simeq 10^{-5}$, as suggested by the upper limit of $\kappa$ in Eq.\eqref{eq: upper limit of kappa}.

\section{Summary and Discussions}
\label{sec:summary}
In this work, we have considered the possibility of realizing the leptogenesis in $U(1)_{L_{\mu}-L_{\tau}}$ symmetric phase. It has been thought that it is hard to realize successful leptogenesis in the $U(1)_{L_{\mu}-L_{\tau}}$ symmetric phase, due to the lack of CP phase in the neutrino mass matrix caused by restriction from the gauge symmetry. However, our investigation reveals a promising avenue for successful leptogenesis by introducing an additional global $U(1)_{B-L}$ symmetry and a scalar field, denoted as $\Phi$, responsible for breaking this symmetry. In the early universe, the rotation of $\Phi$ induces an additional chemical potential in the Boltzmann equation, facilitating successful leptogenesis. In our model, the higher-dimensional term breaking the global $U(1)_{B-L}$ symmetry gives a kick to the phase direction of the scalar $\Phi$ and leads to its rotation in the early universe. This mechanism is called the kinetic misalignment mechanism. Subsequently, as the Hubble friction causes a gradual decrease in the angular component's kinetic energy, rotation halts when the majoron field fails to overcome the potential barrier induced by the higher-dimensional operator term. Following this, the dark matter abundance is generated through the coherent oscillation of the majoron.

We have presented a model capable of predicting neutrino oscillation parameters by leveraging the presence of two zero minors in the mass matrix at low temperatures. This model introduces only three free parameters, namely $\lambda$, $\theta$, and $\phi$, which govern each component of the Majorana mass matrix and determine the characteristic scale of $U(1)_{L_{\mu}-L_{\tau}}$. Through a detailed analysis of these parameters, we observed that for $\lambda \lesssim 8 \times 10^{-7}$, the universe typically resides in the $U(1)_{L_{\mu}-L_{\tau}}$ symmetric phase. To assess the baryon asymmetry, we considered a quartic potential and higher dimensional operator for $\Phi$ and identified regions where both the correct baryon asymmetry and dark matter abundance can be generated. Importantly, this mechanism is expected to be applicable in a broader context, including models with more scalars breaking the gauged $U(1)_{L_{\mu}-L_{\tau}}$ symmetry. This generality arises from the unique determination of the Dirac and Majorana neutrino mass structures in the $U(1)_{L_{\mu}-L_{\tau}}$ symmetric phase, particularly in the context of the type-I seesaw mechanism.

The adoption of a low-scale $U(1)_{L_{\mu}-L_{\tau}}$ symmetry finds motivation in various contexts. As an illustrative example, we focused on a benchmark scenario capable of explaining the muon $g-2$ anomaly through loop corrections involving the $U(1)_{L_{\mu}-L_{\tau}}$ gauge boson $Z'$. Our proposed scenario is particularly effective in the context of leptogenesis when incorporating such a low-scale $U(1)_{L_{\mu}-L_{\tau}}$ symmetry.

\section*{Acknowledgements}
We thank Koichi Hamguchi and Natsumi Nagata for their helpful comments. This work is supported by JSPS KAKENHI Grant (22KJ1050).

\appendix

\section{Derivation of Boltzmann equation in the kinetic motion of majoron background}
\label{appendix: derivation of Boltzmann equation}
In this appendix, we briefly illustrate the deviation of the Boltzmann equation Eq.~\eqref{eq: Boltzmann equation} from the main text. Firstly, we explain how energy level splitting is realized due to the kinetic motion of the majoron background, leading to modified dispersion relations in the case of Dirac and Majorana fermions, as discussed in Refs.~\cite{Ibe:2015nfa, Chun:2023eqc}. Then, we demonstrate the deviation between the thermally averaged interaction rate and the inverse interaction rate in the presence of the kinetic majoron background. Finally, we derive the Boltzmann equation. 

\subsection{Energy level splitting}
The kinetic motion of the majoron background $\dot{\Theta}$ modifies the equations of motion for Dirac and Majorana fermions, resulting in adjusted dispersion relations~\cite{Ibe:2015nfa, Chun:2023eqc}. This is because through this redefinition, we can eliminate the majoron field (or $\Theta$ dependence) in the original Lagrangian, leaving only the new interaction, which is derivative coupling with the majoron: $-\partial_{\mu} \Theta J^{\mu}_{B-L}/2$, where $J^{\mu}_{B-L}$ is the $B-L$ current, as explained in the main text.

In the case of Dirac fermion $\psi_{D}$, the Dirac equation becomes~\cite{Ibe:2015nfa}:
\begin{align}
\label{eq: modified Dirac equation}
\left(i\Slash{\partial}-m-\mu_{\chi}\gamma^0\right)\psi_{D}=0,
\end{align}
where $\mu_{\chi}:=\frac{1}{2}(B-L)_{\psi} \dot{\Theta}$. For $\psi_{D}$ with momentum $p=(E_{D},\bs{p})$, the dispersion relation changes to~\cite{Ibe:2015nfa}:
\begin{align}
E_{D \pm}=\sqrt{|\bs{p}|^2+m^2}\mp \mu_\chi,
\end{align}
due to the non-zero contribution of $\mu_\chi$, contrary to $E_{D}=\sqrt{|\bs{p}|^2+m^2}$ for $\mu_\chi=0$, where the $+$ sign denotes the particle and the $-$ sign denotes the antiparticle. Thus, the term proportional to $\mu_\chi$ causes energy level splittings between leptons and anti-leptons. This fact is crucial when considering the solution of the Dirac equation given by Eq.~\eqref{eq: modified Dirac equation}. Assuming the time-dependence of $\mu_{\chi}$ is negligible, i.e., $\ddot{\Theta}\simeq 0$, we can expand $\psi_{D}$ into plane-wave solutions of the Dirac equation as follows\cite{Ibe:2015nfa}:
\begin{eqnarray}
	& &
	\psi_{D}(x) = \psi_{D,+}(x) + \psi_{D,-}(x),\\
	& &
	\psi_{D,+}(x) = \sum_{s}\int\frac{d^3p}{(2\pi)^3} b_{\bs p,s} u_{\bs p,s} e^{-i \tilde{p}_{D,+} x},~~~
	\psi_{D,-}(x) = \sum_{s}\int\frac{d^3p}{(2\pi)^3}d_{\bs p,s}^\dagger v_{\bs p,s} e^{i \tilde{p}_{D,-} x},
\end{eqnarray}
where $\tilde{p}_{D,+}:=(E_{D +},\bs{p})$ and $\tilde{p}_{D,-}:=(E_{D -},\bs{p})$.
The independent solutions $u_{\bs p,s}$ and $v_{\bs p,s}$ satisfy the Dirac equation for a particle and antiparticle, respectively, and hence, $b_{\bs p,s}$ and $d_{\bs p,s}$ are the creation and annihilation operators of particles and antiparticles. It should be noted that $u_{\bs p,s}$ and $v_{\bs p,s}$ are solutions defined for $\mu_{\chi}=0$. 

In the case of a Majorana fermion $\psi_{M}$, the equation of motion becomes~\cite{Chun:2023eqc}
\begin{align}
\label{eq: modified Majorana equation}
\left(i\bar{\sigma}^{\mu}\partial_{\mu}-\mu_{\chi}\right)\psi_{M}=M \psi_{M}^{\dagger},
\end{align}
where $M$ represents the Majorana mass of $\psi_{M}$ and we adopt a two-component spinor notation here. This equation leads to
\begin{align}
\left(-\partial^2-\mu_{\chi}^2 + 2 i \mu_{\chi} \sigma^j \partial_{j} \right)\psi_{M}=M^2 \psi_{M},
\end{align}
indicating energy level splittings between positive helicity and negative helicity states, denoted as $\psi_{M,\pm}$. For $\psi_{M,\pm}$ with momentum $p=(E_{M\pm},\bs{p})$, the dispersion relation of $\psi_{M\pm}$ is given by~\cite{Chun:2023eqc}
\begin{align}
E_{M \pm}
&=\sqrt{|\bs{p}|^2+M^2 + \mu_{\chi}^2 \mp 2 \mu_{\chi}|\bs{p}| }\nn
&\simeq \sqrt{|\bs{p}|^2+M^2} \mp \mu_{\chi} v(\bs{p}),
\end{align}
where $v(\bs{p})=|\bs{p}|/ \sqrt{|\bs{p}|^2+M^2}$.
Due to the non-zero contribution of $\mu_\chi$, we observe a different dispersion relation, contrary to $E_{M}=\sqrt{M^2+|\bs{p}|^2}$ for $\mu_\chi=0$. Assuming that the time-dependence of $\mu_{\chi}$ is negligible, we can expand $\psi_{M}$ into plane-wave solutions of the equation of motion as follows:
\begin{eqnarray}
	& &
	\psi_{M}(x) = \psi_{M,+}(x) + \psi_{M,-}(x),\\
	& &
	\psi_{M,+}(x) = \int\frac{d^3p}{(2\pi)^3} a_{\bs p,+} x_{\bs p,+} e^{-i \tilde{p}_{M,+} x} + a_{\bs p,+}^\dagger y_{\bs p,+} e^{i \tilde{p}_{M,+} x}\\
    & &
	\psi_{M,-}(x) = \int\frac{d^3p}{(2\pi)^3} a_{\bs p,-} x_{\bs p,-} e^{-i \tilde{p}_{M,-} x} + a_{\bs p,-}^\dagger y_{\bs p,-} e^{i \tilde{p}_{M,-} x},
\end{eqnarray}
where $\tilde{p}_{M,+}:=(E_{M +},\bs{p})$ and $\tilde{p}_{M,-}:=(E_{M -},\bs{p})$. Here, $x_{\bs p,s}$ and $y_{\bs p,s}$ satisfy Dirac equations defined for $\mu_{\chi}=0$, and $a_{\bs p,s}$ and $a_{\bs p,s}^{\dagger}$ are the creation and annihilation operators of this Majorana particle. 

It is crucial to note that the dependence of $\mu_\chi$ only appears in the exponential part of the plane wave, but not in the plane wave solutions that expand fermion fields, i.e., $u_{\bs p,s}$, $v_{\bs p,s}$, $x_{\bs p,s}$, and $y_{\bs p,s}$. This implies that the amplitude, ${\cal M}$, does not depend on $\mu_\chi$, and therefore, the ${\cal S}$ matrix depends on $\mu_\chi$ only through the $\delta$-function of the four-momentum. This consideration is crucial for the subsequent computation.

\subsection{Deviation between thermal-averaged interaction rate and inverse interaction rate}
Next, let us compute the thermal-averaged decay rate and inverse decay rate. Importantly, due to the kinetic motion of the majoron background $\dot{\Theta}$, we observe a difference between the thermal-averaged decay rate and the inverse decay rate. It is noteworthy that the dynamic level splitting discussed in the previous subsection violates CPT-invariance.

First, let's focus on the interaction rate of the right-handed neutrino decaying into the lepton and Higgs, $N_i \to l_\alpha H$, and its inverse decay, $l_\alpha H \to N_i$. We assume that leptogenesis occurs in the $U(1)_{L_{\mu}-L_{\tau}}$ symmetric phase; hence, there is no CP phase in the amplitude $\mathcal{M}$. However, due to the dynamic level splitting, there is a dependence on $p_{\chi}:= (\mu_{\chi},\bs{0})$ in the $\delta$-function of the four-momentum. Therefore, in the presence of the kinetic motion of the majoron background, the thermal-averaged interaction rate of $N_i \to l_\alpha H$ is given by
\begin{align}
&\vev{\Gamma^{\mrm{sym},~\Theta}_{N_i \to l_{\alpha} H}}
=\frac{1}{n_{N_{i}}^{\mrm{eq}}}\int\frac{d^3p_{N_{i}}}{(2\pi)^3 2E_{N_{i}}} 2M_{1} f_{N_{i}}^{\mrm{eq}} ~\sum_{\pm} \Gamma^{\mrm{sym},~\Theta}_{N_{1,\pm} \to l_{\alpha} H}, \\
&\Gamma^{\mrm{sym},~\Theta}_{N_{1,\pm} \to l_{\alpha} H}
:=\frac{1}{2M_{1}} \int\frac{d^3p_{H}}{(2\pi)^3 2E_{H}}\int\frac{d^3p_{l_{\alpha}}}{(2\pi)^3 2E_{l_{\alpha}}} \delta^4(\tilde{p}_{N_{1,\pm}}-\tilde{p}_{l_{\alpha}}-p_{H}) |\mathcal{M} (N_{1,\pm}\to l_\alpha H)|^2
\end{align}
where $\tilde{p}_{N{i,\pm}}:= p_{N_{i}} \mp p_{\chi}v$, $\tilde{p}_{l{\alpha}}:= p_{l_{\alpha}}+ p_{\chi}$, and we approximate the distribution functions by the Maxwell-Boltzmann distribution, $f_{N_{i}}^{\mathrm{eq}} \simeq e^{-E_{N_{i}}/T}$. Similarly, the thermal-averaged interaction rate of $l_\alpha H \to N_{i}$ can be obtained as
\begin{align}
&\vev{\Gamma^{\mrm{sym},~\Theta}_{l_{\alpha} H \to N_{i}}}=\frac{1}{n_{N_{i}}^{\mrm{eq}}}\int\frac{d^3p_{N_{i}}}{(2\pi)^3 2E_{N_{i}}} 2M_{1} f_{N_{i}}^{\mrm{eq}} e^{-\mu_{\chi}/T}  \sum_{\pm} e^{\mp \mu_{\chi}v/T} ~\Gamma^{\mrm{sym},~\Theta}_{N_{1,\pm}\to l_{\alpha} H},
\end{align}
where we have approximated the distribution functions by the Maxwell-Boltzmann distribution, $f_{l_{\alpha}}^{\mathrm{eq}} \simeq e^{-E_{l_{\alpha}}/T},~f_{H}^{\mathrm{eq}}\simeq e^{-E_{H}/T}$, to derive this expression.

Up to the $O(\mu_{\chi})$ order, this reduces to\footnote{
The interaction rate of decay for positive and negative helicity is the same in the absence of kinetic motion of the majoron background, $\Gamma^{\mathrm{sym}}_{N_{1,+} \to l_{\alpha} H}=\Gamma^{\mathrm{sym}}_{N_{1,-} \to l_{\alpha} H}$. Therefore, the $\mu_{\chi}$ dependence from the dynamic level splitting of the right-handed neutrino is canceled out when we sum over the helicity states, up to $O(\mu_{\chi})$:
\begin{align}
\sum_{\pm} e^{\mp \mu_{\chi}v/T} ~\Gamma^{\mrm{sym},~\Theta}_{N_{1,\pm}\to l_{\alpha} H}=\sum_{\pm} ~\Gamma^{\mrm{sym},~\Theta}_{N_{1,\pm}\to l_{\alpha} H} + O(\mu_{\chi}^2).
\end{align}
We note that $\Gamma^{\mathrm{sym},\Theta}_{N_{1,\pm}\to l_{\alpha} H} = \Gamma^{\mathrm{sym}}_{N_{1,\pm}\to l_{\alpha} H} + O(\mu_{\chi})$.
}
\begin{align}
\label{eq: derivation of lepton}
\vev{\Gamma^{\mrm{sym},~\Theta}_{l_{\alpha} H \to N_{i}}} \simeq \vev{\Gamma^{\mrm{sym},~\Theta}_{N_{i} \to l_{\alpha} H}} -\frac{\mu_{\chi}}{T}\vev{\Gamma^{\mrm{sym}}_{N_{i} \to l_{\alpha} H}}.
\end{align}
where $\langle \Gamma^{\mathrm{sym}}_{N_{i} \to l_{\alpha} H} \rangle$ is the thermal-averaged interaction rate of $N_i \to l_\alpha H$ in the absence of the kinetic motion of the majoron background, given in Eq.\eqref{eq: thermal averaged interaction rate}.

Similarly, we can see that
\begin{align}
\label{eq: derivation of anit-lepton}
\vev{\Gamma^{\mrm{sym},~\Theta}_{\bar{l}_{\alpha} \bar{H} \to N_{i}}} \simeq \vev{\Gamma^{\mrm{sym},~\Theta}_{N_{i} \to \bar{l}_{\alpha} \bar{H}}} +\frac{\mu_{\chi}}{T}\vev{\Gamma^{\mrm{sym}}_{N_{i} \to \bar{l}_{\alpha} \bar{H}}}.
\end{align}

Therefore, while the thermal-averaged interaction rate without $\dot{\Theta}$ background follows $\vev{\Gamma^{\mrm{sym}}_{l_{\alpha} H \to N_{i}}} = \vev{\Gamma^{\mrm{sym}}_{N_{i} \to l_{\alpha} H}}$, we have deviation between the thermal-averaged interaction rate and the inverse interaction rate due to the kinetic motion of the majoron background, which induces the non-zero chemical potential for the lepton (and Higgs), as we will explain below.

\subsection{Derivation of the Boltzmann Equation}
Now, we are ready to derive the Boltzmann equation for the lepton asymmetry in the presence of the kinetic motion of the majoron background. We approximate that the distribution function of $X$ is given by $f_X(p)\simeq (n_X/n_X^{\mathrm{eq}})f_X^{\mathrm{eq}}(p)$, assuming kinetic equilibrium.

Furthermore, we approximate $f_X^{\mathrm{eq}}(p)$ by Maxwell-Boltzmann distributions for $X=N$, $l_\alpha$, and $H$ for simplicity. Then, the Boltzmann equations for the leptons and anti-leptons are as follows:\footnote{The number density of positive and negative helicity states of the right-handed neutrino, denoted as $n_{N_{i},\pm}$, should satisfy $\Delta n_{N_{i},\pm}:= n_{N_{i},+}-n_{N_{i},-}=O(\mu_{\chi})$, and $\sum_{\pm} n_{N_{i},\pm}=n_{N_{i}}^{\mathrm{eq}}$, which means that we can express them as $n_{N_{i},\pm} \simeq n_{N_{i}}^{\mathrm{eq}} \pm \Delta n_{N_{i},\pm}/2$. It should be noted that during leptogenesis, right-handed neutrinos are in thermal equilibrium in our model, $n_{N_i}=n_{N_i}^{\mathrm{eq}}$. Using this relation, up to the order of $O(\mu_{\chi})$, we have
\begin{align}
\sum_{\pm}  n_{N_{i},\pm}~\Gamma^{\mrm{sym},~\Theta}_{N_{i,\pm}\to l_{\alpha} H}=n_{N_{i}}^{\mrm{eq}} \sum_{\pm} ~\Gamma^{\mrm{sym},~\Theta}_{N_{i,\pm}\to l_{\alpha} H} + O(\mu_{\chi}^2),
\end{align}
which we use to derive the Boltzmann equations Eqs.\eqref{eq: Boltzmann equation for lepton} and\eqref{eq: Boltzmann equation for anti-lepton}.
}
\begin{align}
\label{eq: Boltzmann equation for lepton}
\dot n_{l_\alpha}+ 3H n_{l_\alpha} 
= &\sum_{i} n^{\mrm{eq}}_{N_i}\vev{\Gamma^{\mrm{sym},~\Theta}_{N_i \to l_{\alpha} H}}
\nn 
&-\frac{n_{l_{\alpha}} n_H}{n_{l_{\alpha}}^{\mrm{eq}} n_H^{\mrm{eq}}}
\sum_{i} n^{\mrm{eq}}_{N_i} \vev{\Gamma^{\mrm{sym},~\Theta}_{l_{\alpha} H \to N_i}},
\\
\label{eq: Boltzmann equation for anti-lepton}
\dot n_{\bar{l}_\alpha}+ 3H n_{\bar{l}_\alpha} 
= &\sum_{i} n^{\mrm{eq}}_{N_i}\vev{\Gamma^{\mrm{sym},~\Theta}_{N_i \to \bar{l}_{\alpha} \bar{H}}},
\nn 
&-\frac{n_{\bar{l}_{\alpha}} n_{\bar{H}}}{n_{\bar{l}_{\alpha}}^{\mrm{eq}} n_{\bar{H}}^{\mrm{eq}}}
\sum_{i} n^{\mrm{eq}}_{N_i} \vev{\Gamma^{\mrm{sym},~\Theta}_{\bar{l}_{\alpha} \bar{H} \to N_i}},
\end{align}
where we have neglected the lepton number violating scattering term because they are a higher order of $\lambda \ll 1$, and decoupled during the leptogenesis.

Then, using the relations obtained in the previous subsection, Eqs.~\eqref{eq: derivation of lepton} and ~\eqref{eq: derivation of anit-lepton}, the Boltzmann equation for the lepton asymmetry density becomes
\begin{align}
\dot{n}_{\Delta l_{\alpha}} + 3 H n_{\Delta l_{\alpha}} = -\sum_{i} n_{N_i}^{\mrm{eq}}\vev{\Gamma^{\mrm{sym}}_{N_i \to l_{\alpha} H}} \left(\frac{n_{\Delta l_\alpha}}{n_{l_\alpha}^{\mrm{eq}}} +\frac{n_{\Delta H}}{n_H^{\mrm{eq}}} -2 \frac{\mu_{\chi}}{T} \right), ~(\alpha = e,\mu, \tau),
\end{align}
where $\mu_{\chi}=\dot{\Theta}/2$.

\bibliographystyle{utphysmod}
\bibliography{ref}

\providecommand{\href}[2]{#2}\begingroup\raggedright\begin{thebibliography}{10}

\bibitem{Fukugita:1986hr}
M.~Fukugita and T.~Yanagida, {\em {Baryogenesis Without Grand Unification}},
  \href{https://dx.doi.org/10.1016/0370-2693(86)91126-3}{Phys.\  Lett.\  B
  {\bfseries 174} (1986) 45--47}.

\bibitem{Kuzmin:1985mm}
V.~A.~Kuzmin, V.~A.~Rubakov, and M.~E.~Shaposhnikov, {\em {On the Anomalous
  Electroweak Baryon Number Nonconservation in the Early Universe}},
  \href{https://dx.doi.org/10.1016/0370-2693(85)91028-7}{Phys.\  Lett.\  B
  {\bfseries 155} (1985) 36}.

\bibitem{Sakharov:1967dj}
A.~D.~Sakharov, {\em {Violation of CP Invariance, C asymmetry, and baryon
  asymmetry of the universe}},
  \href{https://dx.doi.org/10.1070/PU1991v034n05ABEH002497}{Pisma Zh.\  Eksp.\
  Teor.\  Fiz.\  {\bfseries 5} (1967) 32--35}.

\bibitem{Minkowski:1977sc}
P.~Minkowski, {\em {$\mu \to e\gamma$ at a Rate of One Out of $10^{9}$ Muon
  Decays?}},
\href{https://dx.doi.org/10.1016/0370-2693(77)90435-X}{Phys.\  Lett.\
  {\bfseries B67} (1977) 421--428}.

\bibitem{Yanagida:1979as}
T.~Yanagida, {\em {HORIZONTAL SYMMETRY AND MASSES OF NEUTRINOS}},
Conf.\  Proc.\  {\bfseries C7902131} (1979) 95--99.

\bibitem{GellMann:1980vs}
M.~Gell-Mann, P.~Ramond, and R.~Slansky, {\em {Complex Spinors and Unified
  Theories}}, Conf.\  Proc.\  {\bfseries C790927} (1979) 315--321
{\ttfamily [\href{https://arxiv.org/abs/1306.4669}{arXiv:1306.4669}]}.

\bibitem{Mohapatra:1979ia}
R.~N.~Mohapatra and G.~Senjanovic, {\em {Neutrino Mass and Spontaneous Parity
  Violation}},
\href{https://dx.doi.org/10.1103/PhysRevLett.44.912}{Phys.\  Rev.\  Lett.\
  {\bfseries 44} (1980) 912}.

\bibitem{ParticleDataGroup:2022pth}
{\bfseries Particle Data Group} Collaboration, {\em {Review of Particle
  Physics}}, \href{https://dx.doi.org/10.1093/ptep/ptac097}{PTEP {\bfseries
  2022} (2022) 083C01}.

\bibitem{Foot:1990mn}
R.~Foot, {\em {New Physics From Electric Charge Quantization?}},
  \href{https://dx.doi.org/10.1142/S0217732391000543}{Mod.\  Phys.\  Lett.\  A
  {\bfseries 6} (1991) 527--530}.

\bibitem{He:1990pn}
X.~G.~He, G.~C.~Joshi, H.~Lew, and R.~R.~Volkas, {\em {NEW Z-prime
  PHENOMENOLOGY}}, \href{https://dx.doi.org/10.1103/PhysRevD.43.R22}{Physical
  Review D {\bfseries 43} (1991) 22--24}.

\bibitem{He:1991qd}
X.-G.~He, G.~C.~Joshi, H.~Lew, and R.~R.~Volkas, {\em {Simplest Z-prime
  model}}, \href{https://dx.doi.org/10.1103/PhysRevD.44.2118}{Phys.\  Rev.\  D
  {\bfseries 44} (1991) 2118--2132}.

\bibitem{Foot:1994vd}
R.~Foot, X.~G.~He, H.~Lew, and R.~R.~Volkas, {\em {Model for a light Z-prime
  boson}}, \href{https://dx.doi.org/10.1103/PhysRevD.50.4571}{Phys.\  Rev.\  D
  {\bfseries 50} (1994) 4571--4580} {\ttfamily
  [\href{https://arxiv.org/abs/hep-ph/9401250}{hep-ph/9401250}]}.

\bibitem{Branco:1988ex}
G.~C.~Branco, W.~Grimus, and L.~Lavoura, {\em {The Seesaw Mechanism in the
  Presence of a Conserved Lepton Number}},
  \href{https://dx.doi.org/10.1016/0550-3213(89)90304-0}{Nuclear Physics B
  {\bfseries 312} (1989) 492--508}.

\bibitem{Choubey:2004hn}
S.~Choubey and W.~Rodejohann, {\em {A Flavor symmetry for quasi-degenerate
  neutrinos: L(mu) - L(tau)}},
  \href{https://dx.doi.org/10.1140/epjc/s2005-02133-1}{The European Physical
  Journal C {\bfseries 40} (2005) 259--268} {\ttfamily
  [\href{https://arxiv.org/abs/hep-ph/0411190}{hep-ph/0411190}]}.

\bibitem{Araki:2012ip}
T.~Araki, J.~Heeck, and J.~Kubo, {\em {Vanishing Minors in the Neutrino Mass
  Matrix from Abelian Gauge Symmetries}},
  \href{https://dx.doi.org/10.1007/JHEP07(2012)083}{Journal of High Energy
  Physics {\bfseries 07} (2012) 083} {\ttfamily
  [\href{https://arxiv.org/abs/1203.4951}{arXiv:1203.4951}]}.

\bibitem{Heeck:2014sna}
J.~Heeck.
\newblock PhD thesis, Heidelberg U., 2014.

\bibitem{Crivellin:2015lwa}
A.~Crivellin, G.~D'Ambrosio, and J.~Heeck, {\em {Addressing the LHC flavor
  anomalies with horizontal gauge symmetries}},
  \href{https://dx.doi.org/10.1103/PhysRevD.91.075006}{Physical Review D
  {\bfseries 91} (2015) 075006} {\ttfamily
  [\href{https://arxiv.org/abs/1503.03477}{arXiv:1503.03477}]}.

\bibitem{Asai:2017ryy}
K.~Asai, K.~Hamaguchi, and N.~Nagata, {\em {Predictions for the neutrino
  parameters in the minimal gauged U(1)$_{L_\mu-L_\tau}$ model}},
  \href{https://dx.doi.org/10.1140/epjc/s10052-017-5348-x}{Eur.\  Phys.\  J.\
  {\bfseries C77} (2017) 763}
{\ttfamily [\href{https://arxiv.org/abs/1705.00419}{arXiv:1705.00419}]}.

\bibitem{Asai:2018ocx}
K.~Asai, K.~Hamaguchi, N.~Nagata, S.-Y.~Tseng, and K.~Tsumura, {\em {Minimal
  Gauged U(1)$_{L_\alpha - L_\beta}$ Models Driven into a Corner}},
  \href{https://dx.doi.org/10.1103/PhysRevD.99.055029}{Phys.\  Rev.\
  {\bfseries D99} (2019) 055029}
{\ttfamily [\href{https://arxiv.org/abs/1811.07571}{arXiv:1811.07571}]}.

\bibitem{Asai:2019ciz}
K.~Asai, {\em {Predictions for the neutrino parameters in the minimal model
  extended by linear combination of U(1)$_{L_e-L_\mu}$, U(1)$_{L_\mu-L_\tau}$
  and U(1)$_{B-L}$ gauge symmetries}},
  \href{https://dx.doi.org/10.1140/epjc/s10052-020-7622-6}{Eur.\  Phys.\  J.\
  {\bfseries C80} (2020) 76}
{\ttfamily [\href{https://arxiv.org/abs/1907.04042}{arXiv:1907.04042}]}.

\bibitem{Asai:2020qax}
K.~Asai, K.~Hamaguchi, N.~Nagata, and S.-Y.~Tseng, {\em {Leptogenesis in the
  minimal gauged U(1)$_{L_\mu-L_\tau}$ model and the sign of the cosmological
  baryon asymmetry}},
  \href{https://dx.doi.org/10.1088/1475-7516/2020/11/013}{Journal of Cosmology
  and Astroparticle Physics {\bfseries 11} (2020) 013} {\ttfamily
  [\href{https://arxiv.org/abs/2005.01039}{arXiv:2005.01039}]}.

\bibitem{Granelli:2023egb}
A.~Granelli, K.~Hamaguchi, N.~Nagata, M.~E.~Ramirez-Quezada, and J.~Wada, {\em
  {Thermal leptogenesis in the minimal gauged $ \textrm{U}{(1)}_{L_{\mu
  }-{L}_{\tau }} $ model}},
  \href{https://dx.doi.org/10.1007/JHEP09(2023)079}{JHEP {\bfseries 09} (2023)
  079} {\ttfamily [\href{https://arxiv.org/abs/2305.18100}{arXiv:2305.18100}]}.

\bibitem{Gninenko:2001hx}
S.~N.~Gninenko and N.~V.~Krasnikov, {\em {The Muon anomalous magnetic moment
  and a new light gauge boson}},
  \href{https://dx.doi.org/10.1016/S0370-2693(01)00693-1}{Phys.\  Lett.\  B
  {\bfseries 513} (2001) 119} {\ttfamily
  [\href{https://arxiv.org/abs/hep-ph/0102222}{hep-ph/0102222}]}.

\bibitem{Baek:2001kca}
S.~Baek, N.~G.~Deshpande, X.~G.~He, and P.~Ko, {\em {Muon anomalous g-2 and
  gauged L(muon) - L(tau) models}},
  \href{https://dx.doi.org/10.1103/PhysRevD.64.055006}{Phys.\  Rev.\  D
  {\bfseries 64} (2001) 055006} {\ttfamily
  [\href{https://arxiv.org/abs/hep-ph/0104141}{hep-ph/0104141}]}.

\bibitem{Murakami:2001cs}
B.~Murakami, {\em {The Impact of lepton flavor violating Z-prime bosons on muon
  g-2 and other muon observables}},
  \href{https://dx.doi.org/10.1103/PhysRevD.65.055003}{Phys.\  Rev.\  D
  {\bfseries 65} (2002) 055003} {\ttfamily
  [\href{https://arxiv.org/abs/hep-ph/0110095}{hep-ph/0110095}]}.

\bibitem{Ma:2001md}
E.~Ma, D.~P.~Roy, and S.~Roy, {\em {Gauged L(mu) - L(tau) with large muon
  anomalous magnetic moment and the bimaximal mixing of neutrinos}},
  \href{https://dx.doi.org/10.1016/S0370-2693(01)01428-9}{Phys.\  Lett.\  B
  {\bfseries 525} (2002) 101--106} {\ttfamily
  [\href{https://arxiv.org/abs/hep-ph/0110146}{hep-ph/0110146}]}.

\bibitem{Ma:2001tb}
E.~Ma and D.~P.~Roy, {\em {Anomalous neutrino interaction, muon g-2, and atomic
  parity nonconservation}},
  \href{https://dx.doi.org/10.1103/PhysRevD.65.075021}{Phys.\  Rev.\  D
  {\bfseries 65} (2002) 075021} {\ttfamily
  [\href{https://arxiv.org/abs/hep-ph/0111385}{hep-ph/0111385}]}.

\bibitem{Escudero:2019gzq}
M.~Escudero, D.~Hooper, G.~Krnjaic, and M.~Pierre, {\em {Cosmology with A Very
  Light L$_{\mu}$ \ensuremath{-} L$_{\tau}$ Gauge Boson}},
  \href{https://dx.doi.org/10.1007/JHEP03(2019)071}{JHEP {\bfseries 03} (2019)
  071} {\ttfamily [\href{https://arxiv.org/abs/1901.02010}{arXiv:1901.02010}]}.

\bibitem{Araki:2021xdk}
T.~Araki, {\em et al.}, {\em {Resolving the Hubble tension in a
  U(1)$_{L_\mu-L_\tau}$ model with the Majoron}},
  \href{https://dx.doi.org/10.1093/ptep/ptab108}{PTEP {\bfseries 2021} (2021)
  103B05} {\ttfamily
  [\href{https://arxiv.org/abs/2103.07167}{arXiv:2103.07167}]}.

\bibitem{Asai:2021xtg}
K.~Asai, T.~Moroi, and A.~Niki, {\em {Leptophilic Gauge Bosons at ILC Beam Dump
  Experiment}}, \href{https://dx.doi.org/10.1016/j.physletb.2021.136374}{Phys.\
   Lett.\  B {\bfseries 818} (2021) 136374} {\ttfamily
  [\href{https://arxiv.org/abs/2104.00888}{arXiv:2104.00888}]}.

\bibitem{Moroi:2022qwz}
T.~Moroi and A.~Niki, {\em {Leptophilic gauge bosons at lepton beam dump
  experiments}}, \href{https://dx.doi.org/10.1007/JHEP05(2023)016}{JHEP
  {\bfseries 05} (2023) 016} {\ttfamily
  [\href{https://arxiv.org/abs/2205.11766}{arXiv:2205.11766}]}.

\bibitem{Baek:2008nz}
S.~Baek and P.~Ko, {\em {Phenomenology of U(1)(L(mu)-L(tau)) charged dark
  matter at PAMELA and colliders}},
  \href{https://dx.doi.org/10.1088/1475-7516/2009/10/011}{JCAP {\bfseries 10}
  (2009) 011} {\ttfamily
  [\href{https://arxiv.org/abs/0811.1646}{arXiv:0811.1646}]}.

\bibitem{Altmannshofer:2016jzy}
W.~Altmannshofer, S.~Gori, S.~Profumo, and F.~S.~Queiroz, {\em {Explaining dark
  matter and B decay anomalies with an $L_\mu - L_\tau$ model}},
  \href{https://dx.doi.org/10.1007/JHEP12(2016)106}{JHEP {\bfseries 12} (2016)
  106} {\ttfamily [\href{https://arxiv.org/abs/1609.04026}{arXiv:1609.04026}]}.

\bibitem{Arcadi:2018tly}
G.~Arcadi, T.~Hugle, and F.~S.~Queiroz, {\em {The Dark $L_\mu - L_\tau$ Rises
  via Kinetic Mixing}},
  \href{https://dx.doi.org/10.1016/j.physletb.2018.07.028}{Phys.\  Lett.\  B
  {\bfseries 784} (2018) 151--158} {\ttfamily
  [\href{https://arxiv.org/abs/1803.05723}{arXiv:1803.05723}]}.

\bibitem{Bauer:2018egk}
M.~Bauer, S.~Diefenbacher, T.~Plehn, M.~Russell, and D.~A.~Camargo, {\em {Dark
  Matter in Anomaly-Free Gauge Extensions}},
  \href{https://dx.doi.org/10.21468/SciPostPhys.5.4.036}{SciPost Phys.\
  {\bfseries 5} (2018) 036} {\ttfamily
  [\href{https://arxiv.org/abs/1805.01904}{arXiv:1805.01904}]}.

\bibitem{Okada:2019sbb}
N.~Okada and O.~Seto, {\em {Inelastic extra $U(1)$ charged scalar dark
  matter}}, \href{https://dx.doi.org/10.1103/PhysRevD.101.023522}{Phys.\  Rev.\
   D {\bfseries 101} (2020) 023522} {\ttfamily
  [\href{https://arxiv.org/abs/1908.09277}{arXiv:1908.09277}]}.

\bibitem{Borah:2021mri}
D.~Borah, A.~Dasgupta, and D.~Mahanta, {\em {TeV scale resonant leptogenesis
  with L\ensuremath{\mu}-L\ensuremath{\tau} gauge symmetry in light of the muon
  g-2}}, \href{https://dx.doi.org/10.1103/PhysRevD.104.075006}{Phys.\  Rev.\  D
  {\bfseries 104} (2021) 075006} {\ttfamily
  [\href{https://arxiv.org/abs/2106.14410}{arXiv:2106.14410}]}.

\bibitem{Eijima:2023yiw}
S.~Eijima, M.~Ibe, and K.~Murai, {\em {Muon g \ensuremath{-} 2 and non-thermal
  leptogenesis in $ \textrm{U}{(1)}_{L_{\mu }-{L}_{\tau }} $ model}},
  \href{https://dx.doi.org/10.1007/JHEP05(2023)010}{JHEP {\bfseries 05} (2023)
  010} {\ttfamily [\href{https://arxiv.org/abs/2303.09751}{arXiv:2303.09751}]}.

\bibitem{Affleck:1984fy}
I.~Affleck and M.~Dine, {\em {A New Mechanism for Baryogenesis}},
  \href{https://dx.doi.org/10.1016/0550-3213(85)90021-5}{Nucl.\  Phys.\  B
  {\bfseries 249} (1985) 361--380}.

\bibitem{Dine:1995kz}
M.~Dine, L.~Randall, and S.~D.~Thomas, {\em {Baryogenesis from flat directions
  of the supersymmetric standard model}},
  \href{https://dx.doi.org/10.1016/0550-3213(95)00538-2}{Nucl.\  Phys.\  B
  {\bfseries 458} (1996) 291--326} {\ttfamily
  [\href{https://arxiv.org/abs/hep-ph/9507453}{hep-ph/9507453}]}.

\bibitem{Co:2019jts}
R.~T.~Co, L.~J.~Hall, and K.~Harigaya, {\em {Axion Kinetic Misalignment
  Mechanism}}, \href{https://dx.doi.org/10.1103/PhysRevLett.124.251802}{Phys.\
  Rev.\  Lett.\  {\bfseries 124} (2020) 251802} {\ttfamily
  [\href{https://arxiv.org/abs/1910.14152}{arXiv:1910.14152}]}.

\bibitem{Chang:2019tvx}
C.-F.~Chang and Y.~Cui, {\em {New Perspectives on Axion Misalignment
  Mechanism}}, \href{https://dx.doi.org/10.1103/PhysRevD.102.015003}{Phys.\
  Rev.\  D {\bfseries 102} (2020) 015003} {\ttfamily
  [\href{https://arxiv.org/abs/1911.11885}{arXiv:1911.11885}]}.

\bibitem{Co:2019wyp}
R.~T.~Co and K.~Harigaya, {\em {Axiogenesis}},
  \href{https://dx.doi.org/10.1103/PhysRevLett.124.111602}{Phys.\  Rev.\
  Lett.\  {\bfseries 124} (2020) 111602} {\ttfamily
  [\href{https://arxiv.org/abs/1910.02080}{arXiv:1910.02080}]}.

\bibitem{Kusenko:2014uta}
A.~Kusenko, K.~Schmitz, and T.~T.~Yanagida, {\em {Leptogenesis via Axion
  Oscillations after Inflation}},
  \href{https://dx.doi.org/10.1103/PhysRevLett.115.011302}{Phys.\  Rev.\
  Lett.\  {\bfseries 115} (2015) 011302} {\ttfamily
  [\href{https://arxiv.org/abs/1412.2043}{arXiv:1412.2043}]}.

\bibitem{Ibe:2015nfa}
M.~Ibe and K.~Kaneta, {\em {Spontaneous thermal Leptogenesis via Majoron
  oscillation}}, \href{https://dx.doi.org/10.1103/PhysRevD.92.035019}{Phys.\
  Rev.\  D {\bfseries 92} (2015) 035019} {\ttfamily
  [\href{https://arxiv.org/abs/1504.04125}{arXiv:1504.04125}]}.

\bibitem{Domcke:2020kcp}
V.~Domcke, Y.~Ema, K.~Mukaida, and M.~Yamada, {\em {Spontaneous Baryogenesis
  from Axions with Generic Couplings}},
  \href{https://dx.doi.org/10.1007/JHEP08(2020)096}{JHEP {\bfseries 08} (2020)
  096} {\ttfamily [\href{https://arxiv.org/abs/2006.03148}{arXiv:2006.03148}]}.

\bibitem{Co:2020jtv}
R.~T.~Co, N.~Fernandez, A.~Ghalsasi, L.~J.~Hall, and K.~Harigaya, {\em
  {Lepto-Axiogenesis}}, \href{https://dx.doi.org/10.1007/JHEP03(2021)017}{JHEP
  {\bfseries 03} (2021) 017} {\ttfamily
  [\href{https://arxiv.org/abs/2006.05687}{arXiv:2006.05687}]}.

\bibitem{Domcke:2020quw}
V.~Domcke, K.~Kamada, K.~Mukaida, K.~Schmitz, and M.~Yamada, {\em {Wash-In
  Leptogenesis}},
  \href{https://dx.doi.org/10.1103/PhysRevLett.126.201802}{Phys.\  Rev.\
  Lett.\  {\bfseries 126} (2021) 201802} {\ttfamily
  [\href{https://arxiv.org/abs/2011.09347}{arXiv:2011.09347}]}.

\bibitem{Berbig:2023uzs}
M.~Berbig, {\em {Diraxiogenesis}},
  \href{https://dx.doi.org/10.1007/JHEP01(2024)061}{JHEP {\bfseries 01} (2024)
  061} {\ttfamily [\href{https://arxiv.org/abs/2307.14121}{arXiv:2307.14121}]}.

\bibitem{Chao:2023ojl}
W.~Chao and Y.-Q.~Peng, {\em {Majorana Majoron and the Baryon Asymmetry of the
  Universe}}, {\ttfamily
  \href{https://arxiv.org/abs/2311.06469}{arXiv:2311.06469}} (2023).

\bibitem{Chun:2023eqc}
E.~J.~Chun and T.~H.~Jung, {\em {Leptogenesis driven by a Majoron}},
  \href{https://dx.doi.org/10.1103/PhysRevD.109.095004}{Phys.\  Rev.\  D
  {\bfseries 109} (2024) 095004} {\ttfamily
  [\href{https://arxiv.org/abs/2311.09005}{arXiv:2311.09005}]}.

\bibitem{Barnes:2024jap}
P.~Barnes, R.~T.~Co, K.~Harigaya, and A.~Pierce, {\em {Lepto-axiogenesis with
  light right-handed neutrinos}}, {\ttfamily
  \href{https://arxiv.org/abs/2402.10263}{arXiv:2402.10263}} (2024).

\bibitem{Harigaya:2013twa}
K.~Harigaya, T.~Igari, M.~M.~Nojiri, M.~Takeuchi, and K.~Tobe, {\em {Muon g-2
  and LHC phenomenology in the $L_\mu-L_\tau$ gauge symmetric model}},
  \href{https://dx.doi.org/10.1007/JHEP03(2014)105}{JHEP {\bfseries 03} (2014)
  105}
{\ttfamily [\href{https://arxiv.org/abs/1311.0870}{arXiv:1311.0870}]}.

\bibitem{Kaneta:2016uyt}
Y.~Kaneta and T.~Shimomura, {\em {On the possibility of a search for the $L_\mu
  - L_\tau$ gauge boson at Belle-II and neutrino beam experiments}},
  \href{https://dx.doi.org/10.1093/ptep/ptx050}{PTEP {\bfseries 2017} (2017)
  053B04}
{\ttfamily [\href{https://arxiv.org/abs/1701.00156}{arXiv:1701.00156}]}.

\bibitem{Nomura:2019uyz}
T.~Nomura and T.~Shimomura, {\em {Searching for scalar boson decaying into
  light $Z'$ boson at collider experiments in $U(1)_{L_\mu - L_\tau }$ model}},
  \href{https://dx.doi.org/10.1140/epjc/s10052-019-7094-8}{Eur.\  Phys.\  J.\
  {\bfseries C79} (2019) 594}
{\ttfamily [\href{https://arxiv.org/abs/1803.00842}{arXiv:1803.00842}]}.

\bibitem{Dreiner:2008tw}
H.~K.~Dreiner, H.~E.~Haber, and S.~P.~Martin, {\em {Two-component spinor
  techniques and Feynman rules for quantum field theory and supersymmetry}},
  \href{https://dx.doi.org/10.1016/j.physrep.2010.05.002}{Physics Reports
  {\bfseries 494} (2010) 1--196} {\ttfamily
  [\href{https://arxiv.org/abs/0812.1594}{arXiv:0812.1594}]}.

\bibitem{nufit}
{\bfseries NuFIT} Collaboration, {\em NuFIT v5.2}, \url{http://www.nu-fit.org}.

\bibitem{Esteban:2020cvm}
I.~Esteban, M.~C.~Gonzalez-Garcia, M.~Maltoni, T.~Schwetz, and A.~Zhou, {\em
  {The fate of hints: updated global analysis of three-flavor neutrino
  oscillations}}, \href{https://dx.doi.org/10.1007/JHEP09(2020)178}{Journal of
  High Energy Physics {\bfseries 09} (2020) 178} {\ttfamily
  [\href{https://arxiv.org/abs/2007.14792}{arXiv:2007.14792}]}.

\bibitem{LVinZyla:2020zbs}
{J.  Lesgourgues and L.  Verde in P.  A.  Zyla et al.  (Particle Data Group)},
  {\em {Neutrinos in Cosmology (Review of Particle Physics)}},
  \href{https://dx.doi.org/10.1093/ptep/ptaa104}{Progress of Theoretical and
  Experimental Physics {\bfseries 2020} (2020) 083C01}.

\bibitem{Capozzi_2020}
F.~Capozzi, {\em et al.}, {\em {Addendum to ``Global constraints on absolute
  neutrino masses and their ordering''}},
  \href{https://dx.doi.org/10.1103/PhysRevD.101.116013}{Physical Review D
  {\bfseries 101} (2020) 116013}.

\bibitem{Vagnozzi:2017ovm}
S.~Vagnozzi, {\em et al.}, {\em {Unveiling $\nu$ secrets with cosmological
  data: neutrino masses and mass hierarchy}},
  \href{https://dx.doi.org/10.1103/PhysRevD.96.123503}{Phys.\  Rev.\  D
  {\bfseries 96} (2017) 123503} {\ttfamily
  [\href{https://arxiv.org/abs/1701.08172}{arXiv:1701.08172}]}.

\bibitem{Planck:2018vyg}
{\bfseries Planck} Collaboration, {\em {Planck 2018 results. VI. Cosmological
  parameters}},
  \href{https://dx.doi.org/10.1051/0004-6361/201833910}{{Astronomy $\&$
  Astrophysics} {\bfseries 641} (2020) A6} {\ttfamily
  [\href{https://arxiv.org/abs/1807.06209}{arXiv:1807.06209}]}. [Erratum:
  Astron.Astrophys. 652, C4 (2021)].

\bibitem{RoyChoudhury:2019hls}
S.~Roy~Choudhury and S.~Hannestad, {\em {Updated results on neutrino mass and
  mass hierarchy from cosmology with Planck 2018 likelihoods}},
  \href{https://dx.doi.org/10.1088/1475-7516/2020/07/037}{Journal of Cosmology
  and Astroparticle Physics {\bfseries 07} (2020) 037} {\ttfamily
  [\href{https://arxiv.org/abs/1907.12598}{arXiv:1907.12598}]}.

\bibitem{Ivanov:2019hqk}
M.~M.~Ivanov, M.~Simonovi\'c, and M.~Zaldarriaga, {\em {Cosmological Parameters
  and Neutrino Masses from the Final Planck and Full-Shape BOSS Data}},
  \href{https://dx.doi.org/10.1103/PhysRevD.101.083504}{Physical Review D
  {\bfseries 101} (2020) 083504} {\ttfamily
  [\href{https://arxiv.org/abs/1912.08208}{arXiv:1912.08208}]}.

\bibitem{DES:2021wwk}
{\bfseries DES} Collaboration, {\em {Dark Energy Survey Year 3 results:
  Cosmological constraints from galaxy clustering and weak lensing}},
  \href{https://dx.doi.org/10.1103/PhysRevD.105.023520}{Physical Review D
  {\bfseries 105} (2022) 023520} {\ttfamily
  [\href{https://arxiv.org/abs/2105.13549}{arXiv:2105.13549}]}.

\bibitem{Tanseri:2022zfe}
I.~Tanseri, S.~Hagstotz, S.~Vagnozzi, E.~Giusarma, and K.~Freese, {\em {Updated
  neutrino mass constraints from galaxy clustering and CMB lensing-galaxy
  cross-correlation measurements}},
  \href{https://dx.doi.org/10.1016/j.jheap.2022.07.002}{JHEAp {\bfseries 36}
  (2022) 1--26} {\ttfamily
  [\href{https://arxiv.org/abs/2207.01913}{arXiv:2207.01913}]}.

\bibitem{NA64:2024klw}
{\bfseries NA64} Collaboration, {\em {First Results in the Search for Dark
  Sectors at NA64 with the CERN SPS High Energy Muon Beam}},
  \href{https://dx.doi.org/10.1103/PhysRevLett.132.211803}{Phys.\  Rev.\
  Lett.\  {\bfseries 132} (2024) 211803} {\ttfamily
  [\href{https://arxiv.org/abs/2401.01708}{arXiv:2401.01708}]}.

\bibitem{BaBar:2016sci}
{\bfseries BaBar} Collaboration, {\em {Search for a muonic dark force at
  BABAR}}, \href{https://dx.doi.org/10.1103/PhysRevD.94.011102}{Phys.\  Rev.\
  D {\bfseries 94} (2016) 011102} {\ttfamily
  [\href{https://arxiv.org/abs/1606.03501}{arXiv:1606.03501}]}.

\bibitem{CMS:2018yxg}
{\bfseries CMS} Collaboration, {\em {Search for an $L_{\mu}-L_{\tau}$ gauge
  boson using Z$\to4\mu$ events in proton-proton collisions at $\sqrt{s} =$ 13
  TeV}}, \href{https://dx.doi.org/10.1016/j.physletb.2019.01.072}{Phys.\
  Lett.\  B {\bfseries 792} (2019) 345--368} {\ttfamily
  [\href{https://arxiv.org/abs/1808.03684}{arXiv:1808.03684}]}.

\bibitem{Bellini:2011rx}
G.~Bellini {\em et~al.}, {\em {Precision measurement of the 7Be solar neutrino
  interaction rate in Borexino}},
  \href{https://dx.doi.org/10.1103/PhysRevLett.107.141302}{Phys.\  Rev.\
  Lett.\  {\bfseries 107} (2011) 141302} {\ttfamily
  [\href{https://arxiv.org/abs/1104.1816}{arXiv:1104.1816}]}.

\bibitem{CHARM-II:1990dvf}
{\bfseries CHARM-II} Collaboration, {\em {First observation of neutrino trident
  production}}, \href{https://dx.doi.org/10.1016/0370-2693(90)90146-W}{Phys.\
  Lett.\  B {\bfseries 245} (1990) 271--275}.

\bibitem{CCFR:1991lpl}
{\bfseries CCFR} Collaboration, {\em {Neutrino tridents and W Z interference}},
  \href{https://dx.doi.org/10.1103/PhysRevLett.66.3117}{Phys.\  Rev.\  Lett.\
  {\bfseries 66} (1991) 3117--3120}.

\bibitem{Dine:1995uk}
M.~Dine, L.~Randall, and S.~D.~Thomas, {\em {Supersymmetry breaking in the
  early universe}}, \href{https://dx.doi.org/10.1103/PhysRevLett.75.398}{Phys.\
   Rev.\  Lett.\  {\bfseries 75} (1995) 398--401} {\ttfamily
  [\href{https://arxiv.org/abs/hep-ph/9503303}{hep-ph/9503303}]}.

\bibitem{Planck:2018jri}
{\bfseries Planck} Collaboration, {\em {Planck 2018 results. X. Constraints on
  inflation}}, \href{https://dx.doi.org/10.1051/0004-6361/201833887}{Astron.\
  Astrophys.\  {\bfseries 641} (2020) A10} {\ttfamily
  [\href{https://arxiv.org/abs/1807.06211}{arXiv:1807.06211}]}.

\bibitem{Kawasaki:2006yb}
M.~Kawasaki and K.~Nakayama, {\em {Affleck-Dine baryogenesis in
  anomaly-mediated SUSY breaking}},
  \href{https://dx.doi.org/10.1088/1475-7516/2007/02/002}{JCAP {\bfseries 02}
  (2007) 002} {\ttfamily
  [\href{https://arxiv.org/abs/hep-ph/0611320}{hep-ph/0611320}]}.

\bibitem{Enomoto:2023sma}
K.~Enomoto, K.~Hamaguchi, K.~Kamada, and J.~Wada, {\em {Revisiting Affleck-Dine
  leptogenesis with light sleptons}},
  \href{https://dx.doi.org/10.1088/1475-7516/2023/07/003}{JCAP {\bfseries 07}
  (2023) 003} {\ttfamily
  [\href{https://arxiv.org/abs/2304.05614}{arXiv:2304.05614}]}.

\bibitem{Barbieri:1999ma}
R.~Barbieri, P.~Creminelli, A.~Strumia, and N.~Tetradis, {\em {Baryogenesis
  through leptogenesis}},
  \href{https://dx.doi.org/10.1016/S0550-3213(00)00011-0}{Nucl.\  Phys.\  B
  {\bfseries 575} (2000) 61--77} {\ttfamily
  [\href{https://arxiv.org/abs/hep-ph/9911315}{hep-ph/9911315}]}.

\bibitem{Engelhard:2006yg}
G.~Engelhard, Y.~Grossman, E.~Nardi, and Y.~Nir, {\em {The Importance of N2
  leptogenesis}},
  \href{https://dx.doi.org/10.1103/PhysRevLett.99.081802}{Phys.\  Rev.\  Lett.\
   {\bfseries 99} (2007) 081802} {\ttfamily
  [\href{https://arxiv.org/abs/hep-ph/0612187}{hep-ph/0612187}]}.

\bibitem{Antusch:2010ms}
S.~Antusch, P.~Di~Bari, D.~A.~Jones, and S.~F.~King, {\em {A fuller flavour
  treatment of $N_2$-dominated leptogenesis}},
  \href{https://dx.doi.org/10.1016/j.nuclphysb.2011.10.036}{Nucl.\  Phys.\  B
  {\bfseries 856} (2012) 180--209} {\ttfamily
  [\href{https://arxiv.org/abs/1003.5132}{arXiv:1003.5132}]}.

\bibitem{Blanchet:2011xq}
S.~Blanchet, P.~Di~Bari, D.~A.~Jones, and L.~Marzola, {\em {Leptogenesis with
  heavy neutrino flavours: from density matrix to Boltzmann equations}},
  \href{https://dx.doi.org/10.1088/1475-7516/2013/01/041}{JCAP {\bfseries 01}
  (2013) 041} {\ttfamily
  [\href{https://arxiv.org/abs/1112.4528}{arXiv:1112.4528}]}.

\bibitem{Abada:2006fw}
A.~Abada, S.~Davidson, F.-X.~Josse-Michaux, M.~Losada, and A.~Riotto, {\em
  {Flavor issues in leptogenesis}},
  \href{https://dx.doi.org/10.1088/1475-7516/2006/04/004}{JCAP {\bfseries 04}
  (2006) 004} {\ttfamily
  [\href{https://arxiv.org/abs/hep-ph/0601083}{hep-ph/0601083}]}.

\bibitem{Nardi:2006fx}
E.~Nardi, Y.~Nir, E.~Roulet, and J.~Racker, {\em {The Importance of flavor in
  leptogenesis}}, \href{https://dx.doi.org/10.1088/1126-6708/2006/01/164}{JHEP
  {\bfseries 01} (2006) 164} {\ttfamily
  [\href{https://arxiv.org/abs/hep-ph/0601084}{hep-ph/0601084}]}.

\bibitem{Buchmuller:2004nz}
W.~Buchmuller, P.~Di~Bari, and M.~Plumacher, {\em {Leptogenesis for
  pedestrians}}, \href{https://dx.doi.org/10.1016/j.aop.2004.02.003}{Annals
  Phys.\  {\bfseries 315} (2005) 305--351} {\ttfamily
  [\href{https://arxiv.org/abs/hep-ph/0401240}{hep-ph/0401240}]}.

\bibitem{Preskill:1982cy}
J.~Preskill, M.~B.~Wise, and F.~Wilczek, {\em {Cosmology of the Invisible
  Axion}}, \href{https://dx.doi.org/10.1016/0370-2693(83)90637-8}{Phys.\
  Lett.\  B {\bfseries 120} (1983) 127--132}.

\bibitem{Abbott:1982af}
L.~F.~Abbott and P.~Sikivie, {\em {A Cosmological Bound on the Invisible
  Axion}}, \href{https://dx.doi.org/10.1016/0370-2693(83)90638-X}{Phys.\
  Lett.\  B {\bfseries 120} (1983) 133--136}.

\bibitem{Dine:1982ah}
M.~Dine and W.~Fischler, {\em {The Not So Harmless Axion}},
  \href{https://dx.doi.org/10.1016/0370-2693(83)90639-1}{Phys.\  Lett.\  B
  {\bfseries 120} (1983) 137--141}.

\bibitem{Audren:2014bca}
B.~Audren, J.~Lesgourgues, G.~Mangano, P.~D.~Serpico, and T.~Tram, {\em
  {Strongest model-independent bound on the lifetime of Dark Matter}},
  \href{https://dx.doi.org/10.1088/1475-7516/2014/12/028}{JCAP {\bfseries 12}
  (2014) 028} {\ttfamily
  [\href{https://arxiv.org/abs/1407.2418}{arXiv:1407.2418}]}.

\bibitem{Enqvist:2019tsa}
K.~Enqvist, S.~Nadathur, T.~Sekiguchi, and T.~Takahashi, {\em {Constraints on
  decaying dark matter from weak lensing and cluster counts}},
  \href{https://dx.doi.org/10.1088/1475-7516/2020/04/015}{JCAP {\bfseries 04}
  (2020) 015} {\ttfamily
  [\href{https://arxiv.org/abs/1906.09112}{arXiv:1906.09112}]}.

\bibitem{Nygaard:2020sow}
A.~Nygaard, T.~Tram, and S.~Hannestad, {\em {Updated constraints on decaying
  cold dark matter}},
  \href{https://dx.doi.org/10.1088/1475-7516/2021/05/017}{JCAP {\bfseries 05}
  (2021) 017} {\ttfamily
  [\href{https://arxiv.org/abs/2011.01632}{arXiv:2011.01632}]}.

\bibitem{Alvi:2022aam}
S.~Alvi, T.~Brinckmann, M.~Gerbino, M.~Lattanzi, and L.~Pagano, {\em {Do you
  smell something decaying? Updated linear constraints on decaying dark matter
  scenarios}}, \href{https://dx.doi.org/10.1088/1475-7516/2022/11/015}{JCAP
  {\bfseries 11} (2022) 015} {\ttfamily
  [\href{https://arxiv.org/abs/2205.05636}{arXiv:2205.05636}]}.

\bibitem{Simon:2022ftd}
T.~Simon, G.~Franco~Abell\'an, P.~Du, V.~Poulin, and Y.~Tsai, {\em
  {Constraining decaying dark matter with BOSS data and the effective field
  theory of large-scale structures}},
  \href{https://dx.doi.org/10.1103/PhysRevD.106.023516}{Phys.\  Rev.\  D
  {\bfseries 106} (2022) 023516} {\ttfamily
  [\href{https://arxiv.org/abs/2203.07440}{arXiv:2203.07440}]}.

\end{thebibliography}\endgroup

\end{document}